\documentclass[preprint,mathptmx,jeeabib]{jeea}
\let\phi\nonvarphi
\usepackage[french,italian,american]{babel}
\usepackage{amsmath,amssymb,amsfonts}
\usepackage{graphicx}
\usepackage{tikz}
\usetikzlibrary{trees,shapes,backgrounds}
\usepackage{comment}
\usepackage{pdfpages}
\usepackage{xr}
\usepackage{ubar}
\DeclareMathOperator{\interior}{int}

\newcommand\xqed[1]{%
  \leavevmode\unskip\penalty9999 \hbox{}\nobreak\hfill
  \quad\hbox{#1}}
\newcommand\demo{\xqed{$\triangle$}}

\definecolor{ballblue}{rgb}{0.13, 0.67, 0.8}

\usepackage[colorlinks = true,
            linkcolor = blue,
            urlcolor  = blue,
            citecolor = blue,
            anchorcolor = blue]{hyperref}

\Title{Recursive Preferences, Correlation Aversion, and the Temporal Resolution of Uncertainty}
\Runninghead{Stanca}{Recursive Preferences and Temporal Resolution}
\Author{Lorenzo}{Stanca}{ESOMAS Department and Collegio Carlo Alberto, Università di Torino, Corso Unione Sovietica, 218 Bis 10134, Turin, Italy.}{lorenzomaria.stanca@unito.it}
\Thanks{This paper is a revised and extended version of Chapter 1 of my dissertation at Northwestern University; several results also appeared in my job market paper which shared the same title. I thank four anonymous referees for their constructive comments and suggestions. I am grateful to Nabil Al-Najjar, Peter Klibanoff, and Marciano Siniscalchi for their help and guidance. I thank Antoine Bommier,  Francesco Fabbri, Veronica Merlone, Ignacio Monzón, Daniele Pennesi, and Giulio Principi for their feedback.}
\Abstract{This paper investigates a novel behavioral feature of recursive preferences: aversion to risks that persist over time, or simply \textit{correlation aversion}. Greater persistence provides information about future consumption but reduces opportunities to hedge consumption risk. I show that, for recursive preferences that exhibit a  preference for early resolution of uncertainty, correlation aversion is equivalent to increasing relative risk aversion.  To quantify correlation aversion, I develop the concept of the persistence premium, which measures how much an individual is willing to pay to eliminate persistence in consumption.  I provide an approximation of the persistence premium in the spirit of Arrow--Pratt, which provides a quantitative representation of the trade-off between information and hedging.  I show that correlation-averse preferences have a variational representation, linking correlation aversion to concerns about model misspecification. I present several applications. I first illustrate how correlation aversion shapes portfolio choices, and then show how the persistence premium can improve the calibration of macro-finance models. In an optimal taxation model, I show that recursive preferences---unlike standard preferences---lead to redistributive tax policies that increase social mobility.}

\begin{document}

\nocite{de1962sul}
\nocite{savage1954foundations}

\maketitle

\section{Introduction}

Recursive preferences are of central importance in many economic applications, including models of consumption-based asset pricing  \citep{epstein1989substitution, epstein1991substitution}, precautionary savings \citep{weil1989equity, hansen1999robust}, business cycles \citep{tallarini2000risk},  progressive taxation and inequality \citep{benabou2002tax}, and risk-sharing \citep{epstein2001sharing, anderson2005dynamics}. Recursive preferences have also been applied to   climate change \citep{bansal2017climate, cai2019social},  optimal fiscal policy \citep{karantounias2018optimal}, and repeated games \citep{kochov2016repeated}.


A key feature of recursive preferences is their ability to distinguish between risk aversion and intertemporal substitution---two important preference parameters that, for both theoretical and empirical reasons, should  be disentangled. In this paper, I show that recursive preferences also exhibit sensitivity to a third behavioral trait: \textit{aversion to risks that persist over time}. This trait, which has received less attention in the past, plays a significant role in many economic applications.

I introduce a new axiom, which I call \textit{correlation aversion}, that  captures a preference for avoiding risks that are correlated over time. Greater persistence in risks reduces the ability to hedge, which   a  risk averse decision maker (DM)  dislikes. However, increased correlation  also means greater informativeness about future consumption.  This fact creates a  trade-off between \textit{hedging} and \textit{information}, which is central to this paper.

To illustrate, compare two gambles: in gamble \(A\), a single coin flip at \(t=1\) determines all future consumption (all $1$’s or all $0$’s), whereas in gamble \(B\) a fair coin is tossed each period (yielding $1$ or $0$ each time). A hedging motive suggests a preference for $B$ over $A$, but $A$ resolves all risk at $t=1$, providing more non-instrumental information about future consumption.\footnote{Here I refer to risk concerning consumption, not income; hence, early coin tosses offer no planning advantage---information is non-instrumental.}  \cite{kreps1978temporal} show that recursive preferences imply a preference for early resolution of uncertainty, or equivalently, non-instrumental information. Hence, the comparison is not straightforward: $A$ resolves risk early, while $B$ is preferred for its hedging value. My notion of correlation aversion requires the hedging motive to outweigh the preference for non-instrumental information, leading a DM to prefer $B$ over $A$.  
 
Correlation aversion, therefore, should presumably be  reflected in a DM's preferences through a limited willingness to pay for \textit{non-instrumental information} about future consumption. I characterize correlation aversion for recursive preferences that exhibit a preference for early resolution of uncertainty,  showing that  it is equivalent to increasing relative risk aversion (IRRA).   Consistent with the intuition, IRRA  imposes bounds on the demand for non-instrumental information, and a mild strengthening of IRRA guarantees that recursive preferences admit a representation reflecting robustness to model misspecification.

Further, I introduce the \textit{persistence premium}, a measure of correlation aversion that reflects the willingness to pay to eliminate risk persistence. I derive an  approximation  of this premium in the spirit of Arrow–Pratt that links correlation aversion to  preference parameters such as  risk aversion and preference for non-instrumental information.

I then  apply these results to asset pricing and  income taxation. I illustrate how correlation aversion  shapes portfolio choices and asset prices and how the persistence premium improves macro-finance model calibrations. Finally, I show how correlation-averse  preferences shape progressive tax structures by increasing redistribution in a way that promotes social mobility.


\vspace*{-3mm}
\subsection*{Preview of results}
\vspace*{-3mm} 
I consider      recursive preferences    characterized by three components \((\phi,u,\beta)\): \(\phi\) reflects risk attitudes, \(u\) determines elasticity of intertemporal substitution (EIS), and \(\beta\) captures time preference.

First, I reframe preferences for early resolution of uncertainty in terms of the Blackwell order of informativeness \citep{blackwell1951comparison}. A preference for  early resolution of uncertainty implies the DM prefers more informative lotteries (see  the discussion after  Definition \ref{def:peru}). Propositions \ref{peruchar} and \ref{prop:upidara} connect    preferences for early resolution of uncertainty with decreasing absolute risk aversion.

I then use this result to  introduce a   measure of attitudes toward early resolution of uncertainty, which I refer to as $ER_\phi$  (see equation \ref{equation:er}  and Appendix \ref{subsec:measuringperu}). I develop a foundation of  this measure  by providing  an approximation of the early resolution premium, which quantifies the willingness to pay to have uncertainty resolve  early. This approximation shows that the premium depends positively on $ER_\phi$ (see Corollary \ref{corollaryperu}).

Next, I introduce a novel definition of correlation aversion. More correlation adds informativeness in the Blackwell sense (Proposition \ref{iidinfo}), creating conflicting incentives for a DM with recursive preferences, who prefers less persistent lotteries but values the  information. This result formalizes the key trade-off explored in this paper  between intertemporal hedging and non-instrumental information, as illustrated in the initial example: \( A \) is more informative than \( B \), but \( B \) provides better hedging value.

The main result, Theorem \ref{theo1}, shows that for a DM who prefers early resolution for every possible value of $\beta$, correlation aversion is equivalent to $\phi$ satisfying IRRA.   I further show that IRRA  limits how much a DM values non-instrumental information (see equation \ref{eq:boundperu}).  Notably, IRRA encompasses common recursive utility models like Epstein–Zin.

To measure correlation aversion, I introduce the \textit{persistence premium}, which quantifies how much a DM is willing to pay to eliminate  consumption persistence (see equation \ref{eq:corrpremium}). Using an approximation à la Arrow–Pratt I find a formula  that connects
 correlation aversion with  risk aversion, EIS, persistence, and preference for information.  
 
 To illustrate, in the Epstein–Zin case—where \(1-\alpha\) is the coefficient of relative risk aversion, \({1}/{(1-\rho)}\) is the elasticity of intertemporal substitution, and \(\varepsilon\in[0,1]\) measures consumption persistence---the premium is approximately given by
\[
\tilde{a} + \tilde{b}\,\varepsilon \Bigl(1 - \frac{\alpha}{\rho}\Bigr)\Bigl(\frac{1}{c_H} + \frac{1}{c_L}\Bigr)
\;-\;
\tilde{c}\,(\varepsilon^2 - 1)\,ER_{\phi}\;,
\]
where \(\tilde{a}, \tilde{b}, \tilde{c} > 0\) and consumption can be either high (\(c_H\)) or low (\(c_L\)). Hence, the premium rises with consumption persistence but at a decreasing rate. In particular, higher risk aversion makes the premium increase more rapidly as \(\varepsilon\) grows, while a higher \(ER_{\phi}\) moderates this increase. Under IRRA, this result generalizes (see Corollary \ref{premiumcorollary} and the related discussion), formalizing the quantitative trade-off between information and hedging.

To further clarify the role of risk attitudes, Theorem \ref{theo2} shows that a mild strengthening of IRRA implies that recursive preferences admit a representation reflecting robustness to model misspecification---a concern that future consumption distributions may be wrong. This result extends earlier work (e.g., \cite{hansen1999robust}) that links multiplier preferences to robustness, generalizing it to  all correlation-averse preferences.

I then  provide applications of these results.

\vspace{0.15cm}
\noindent
\textbf{Asset pricing}.  These results imply that correlation aversion significantly influences portfolio choices at both individual investor and macroeconomic levels. At the investor level, I discuss how risk preferences affect investment strategies. All other things being equal, investors characterized by high relative risk aversion and a low preference for information are more likely to favor bonds over stocks. Conversely, investors who exhibit a strong preference for information relative to their degree of risk aversion will find stocks more attractive due to the news they provide about  long-run consumption growth.  

At the macro-financial level, correlation aversion helps explain the equity premium puzzle—the high observed excess returns investors require for holding equities.  The long-run risk model of \cite{bansal2004risks}, combined with Epstein–Zin preferences and persistent  consumption growth, successfully matches the observed equity premium.   Theorem \ref{theo1} implies that   the equity premium is higher under Epstein–Zin preferences because they exhibit IRRA and therefore correlation aversion, but  \textit{despite} their preference for non-instrumental information.

 This analysis highlights the need to calibrate preference parameters to achieve a reasonable level of correlation aversion. Drawing  inspiration from \cite{epstein2014much}, I introduce an analogue of the persistence premium in this macro-finance setting  which asks: ``What fraction of your wealth would you sacrifice to eliminate all persistence in consumption growth?''  Using existing experimental evidence, I show that---under standard parameters commonly used in the literature---the persistence premium proves unreasonably high (see Section \ref{impliforassetpricing}).

This issue  arises because Epstein–Zin preferences do not fully disentangle risk aversion from preferences for early resolution. To address this evidence, I explore a generalization of Epstein–Zin preferences to hyperbolic absolute  risk aversion (HARA), which partially separates these preference parameters (see equation  \eqref{def:haraprefs}). I show that these preferences can exhibit a level of correlation aversion comparable to the standard Epstein–Zin parametrization used in the literature, but with a lower level of risk aversion that is more consistent with empirical evidence. Hence,  this new model suggests that one can explain investors'  preference for bonds over stocks without assuming unrealistically high risk aversion.

\vspace{0.15cm}

\noindent
\textbf{Income  taxation and social mobility}. Progressive income taxation is often viewed as a  key tool for addressing income inequality \citep{diamond2011case}.  I show that correlation aversion introduces social mobility as an additional motive for progressive taxation. Under recursive utility with correlation aversion, I obtain a normative foundation for dynamic redistribution  that takes the form of an ``inheritance'' tax: redistribution goes   from historically high human capital households to historically low human capital ones. Correlation aversion would also favor other policies that target persistent inequalities rather than just smoothing short-term shocks---such as redistributive education financing.

I consider a simplified version of \citeauthor{benabou2002tax}'s (\citeyear{benabou2002tax}) stochastic model of human capital accumulation. While standard discounted expected utility implies that the optimal level of progressive taxation is largely unaffected by human capital persistence, recursive utility with correlation aversion implies that greater persistence significantly increases     progressivity of the optimal income tax. Consequently, social mobility is higher under correlation aversion (see Section \ref{app:progtax}).

The intuition is that higher inheritability of human capital increases consumption persistence: individuals from historically high human-capital families tend to have higher incomes and consumption, and vice versa. Correlation aversion therefore favors reducing this persistence.

\vspace*{-3mm}
\subsection*{Related literature}
\vspace*{-3mm}

The theoretical literature on dynamic choice has considered a notion of correlation aversion derived from the literature on risk aversion with multiple commodities started  by \cite{kihlstrom1974risk} (see also \citealt{definetti1952sulla}, \citealt{richard1975multivariate},  \citealt{epstein1980increasing}, and \citealt{finkelshtain1999risk}).   In particular, \cite{bommier2007risk} considers a notion of correlation aversion based on the \citeauthor{kihlstrom1974risk} approach in a continuous time setting. \cite{kochov2015time} and \cite{bommier2019ambiguity} study the extension to a purely subjective setting of this property, which they refer to as intertemporal hedging. 

 Intertemporal hedging involves comparing intertemporal gambles that do not differ in terms of temporal resolution of uncertainty. \cite{miao2015risk} and \cite{andersen2018multiattribute} relate Epstein–Zin utility to an analogous notion of intertemporal hedging and provide experimental evidence in its favor. I show that within the class of recursive preferences identified by $(\phi,u,\beta)$---which I refer to as Kreps-Porteus (KP) preferences---intertemporal hedging is equivalent to $\phi$ being concave, i.e., risk aversion (see Section \ref{negativecorrelation}).
  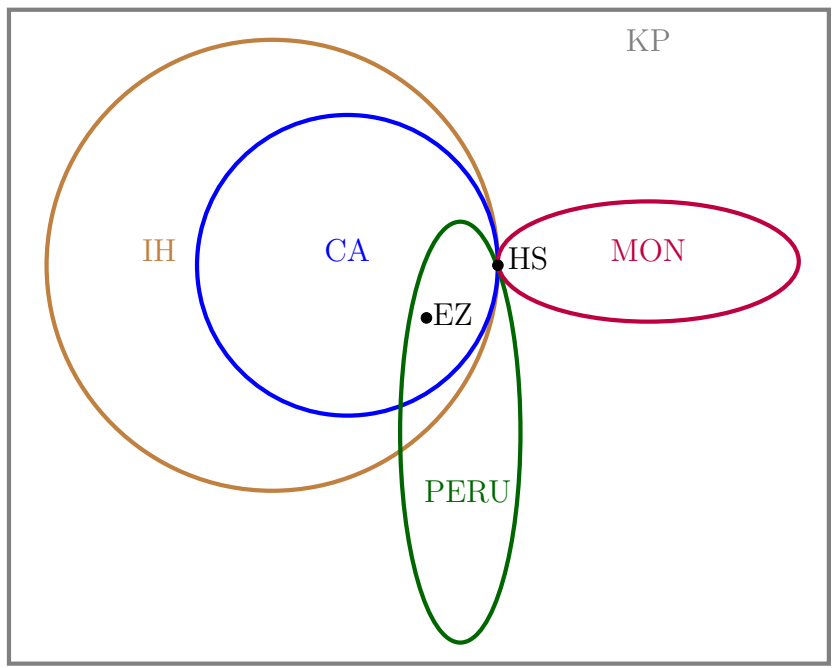
\begin{figure}[!ht]
\begin{center}
\begin{tikzpicture}
\filldraw[gray,ultra thick,fill=white] (-3,-4.3) rectangle (7.9,4.4);
\scope 
\clip (0,0) circle (0);
\fill[white] (1,0) circle (1);
\endscope
\draw[brown,ultra thick] (0.5,1) circle (3);
 \draw[color=blue!100,ultra thick]     (1.1,1) circle (2.4);
  \draw[color=ballblue!100,ultra thick]     (2.488,0.8) circle (1);
  \draw[color=black!60!green,ultra thick]    (2.2,-0.5) ellipse (1.5cm and 3cm);
  \draw[purple,ultra thick]    (5.5,1.05) ellipse (2cm and 0.8cm);
      \draw[color=black!60!green,ultra thick] (2.8,-2) node[] { \text{PERU}};
      \draw[color=blue!100,ultra thick]   (0,1.2) node[] { \text{CA}};
      \draw[color=ballblue!100,ultra thick]   (2.2,1.2) node[] { \text{SCA}};
      \draw[brown] (-1.7,1.2) node[] { \text{IH}};
      \draw[purple,ultra thick] (5.5,1.2) node[] { \text{MON}};
      \draw (3.9,1.1) node[ultra thick] { \text{HS}};
       \draw[gray,ultra thick] (5.5,4) node[] { \text{KP}};
       \draw[fill=black] (3.5,1) circle (0.07);
       \draw[fill=black] (2.55,0.30) circle (0.07);
        \draw (2.9,0.35) node[ultra thick] {  \text{EZ}};
\end{tikzpicture}
\end{center}
\caption{Relationship between correlation-averse (CA) preferences and other recursive (KP)  preferences: recursive preferences that satisfy intertemporal-hedging (IH), Epstein–Zin (EZ) preferences, 
multiplier-preferences (HS), monotone recursive preferences (MON),  preferences that exhibit a preference for early resolution of uncertainty (PERU), and strong correlation aversion (SCA). HS preferences are the only ones that exhibit all these features at the same time.}\label{figure1}
\end{figure}

  
I also consider the notion of strong correlation aversion, which strengthens IRRA and implies a robust representation of preferences.      \citeauthor{epstein1989substitution}'s (\citeyear{epstein1989substitution}) preferences  and \citeauthor{hansen2001robust}'s  (\citeyear{hansen2001robust}) multiplier preferences satisfy this condition. \cite{meyer2019generalized} first provided  a connection between Epstein–Zin preferences and model misspecification. Within the Kreps-Porteus setting, multiplier preferences are the only ones to jointly satisfy strong correlation aversion, preference for early resolution of uncertainty, and  monotonicity as defined in \cite{bommier2017monotone}. Figure \ref{figure1} illustrates the relationship just discussed between correlation aversion  and other prominent classes of recursive preferences.  
 I discuss the relationship of correlation aversion with the work of \cite{dejarnette2020time} and \cite{dillenberger2020stochastic} on preferences that satisfy stochastic impatience   in more depth in Section \ref{implications}.  

Similarly to \cite{andreasen2020importance}, I introduce a generalization of Epstein–Zin preferences in order to disentangle   risk aversion from attitudes toward non-instrumental information. Their generalization is able to resolve puzzles in the long-run risk model.  The main difference with my approach is that they propose a more general form for the utility function $u$, while I propose a more general formulation of  $\phi$.  

\cite{grant1998intrinsic} also provide a connection between preference for non-instrumental information and the Blackwell order in a setting in which preferences are defined over two-stage lotteries  (see also \citealt{dillenberger2010preferences}). In their framework, each information system induces a two-stage lottery. In contrast, in the present setting with temporal lotteries, consumption in one period serves as a signal for information in the subsequent period.
\vspace{0.15cm}

\section{Preliminaries}\label{prelim}
\noindent
\textbf{Choice setting}. I assume that time is discrete and varies over a finite horizon \(2 \leq T < \infty\). The Supplemental Appendix (see Section \ref{infinite}) describes the setting for an infinite horizon, i.e., \(T = \infty\). I assume that the consumption set \(C\) satisfies either \(C = [0, \infty)\) or \(C = (0, \infty)\), depending on the specific recursive representation under consideration.  Given a Polish space $X$, let $\Delta_s(X),\Delta_b(X)$ denote the space of simple (i.e., finite support) and Borel probability measures with bounded support over $X$, respectively.  Observe that $\Delta_s(X)\subseteq\Delta_b(X)$, and  that both  are convex spaces.
 
 Given $\ell, m \in \Delta_b(X)$ such that $\ell$ is absolutely continuous with respect to $m$ (denoted by $\ell \ll m$), I denote the Radon-Nikodym derivative by $\frac{d\ell}{dm}$. For $x\in X$, let $\delta_x \in \Delta_b(X)$ represent the Dirac probability, defined by $\delta_x(A) = 1$ when $x \in A$ and $\delta_x(A) = 0$ when $x \notin A$.   I denote by  $\bigoplus_{i=1}^n \pi_i m_i$  the convex combination of $n$ probabilities $\left(m_i\right)_{i=1}^{n}$ in $\Delta_b(X)$ with a probability vector $\left(\pi_i\right)_{1 \leq i \leq n}$. Note that every two-stage lottery $m\in\Delta_s(\Delta_s(X))$ can be   associated  to a matrix-vector pair $(M[m],\mu[m])$ where  $M[m]$ is a     stochastic matrix whose rows describe each probability $M[m](\cdot|i)\in \mathrm{supp}m$ in the support of $m$ for $i=1,\ldots,|\mathrm{supp}m|$, and   $\mu[m]$ is a  probability row vector which describes the probability of each $i$.

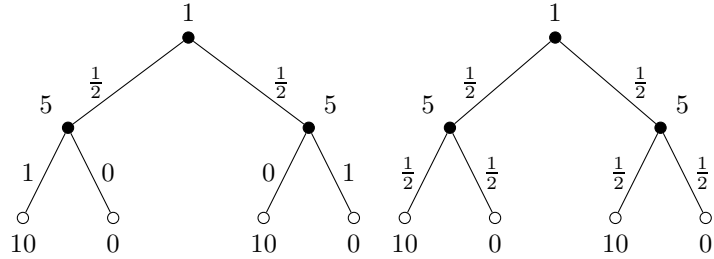
\begin{figure}
\begin{center}
\begin{tikzpicture}[scale=0.8,font=\footnotesize]
\tikzstyle{solid node}=[circle,draw,inner sep=1.5,fill=black]
\tikzstyle{hollow node}=[circle,draw,inner sep=1.5]
\tikzstyle{level 1}=[level distance=15mm,sibling distance=4cm]
\tikzstyle{level 2}=[level distance=15mm,sibling distance=1.5cm]
\tikzstyle{level 3}=[level distance=15mm,sibling distance=1cm]
\node(0)[solid node,label=above:{$1$}]{}
child{node[solid node,label=above left:{$5$}]{}
child{node[hollow node,label=below:{$10$}]{} edge from parent node[left]{$	1$}}
child{node[hollow node,label=below:{$0$}]{} edge from parent node[right]{$0$}}
edge from parent node[left,xshift=-5]{$\frac{1}{2}$}
}
child{node[solid node,label=above right:{$5$}]{}
child{node[hollow node,label=below:{$10$}]{} edge from parent node[left]{$0$}}
child{node[hollow node,label=below:{$0$}]{} edge from parent node[right]{$1$}}
edge from parent node[right,xshift=5]{$\frac{1}{2}$}
};
\end{tikzpicture}
\begin{tikzpicture}[scale=0.8,font=\footnotesize]
\tikzstyle{solid node}=[circle,draw,inner sep=1.5,fill=black]
\tikzstyle{hollow node}=[circle,draw,inner sep=1.5]
\tikzstyle{level 1}=[level distance=15mm,sibling distance=3.5cm]
\tikzstyle{level 2}=[level distance=15mm,sibling distance=1.5cm]
\tikzstyle{level 3}=[level distance=15mm,sibling distance=1cm]
\node(0)[solid node,label=above:{$1$}]{} 
child{node[solid node,label=above left:{$5$}]{} 
child{node[hollow node,label=below:{$10$}]{} edge from parent node[left]{$\frac{1}{2}$}}
child{node[hollow node,label=below:{$0$}]{} edge from parent node[right]{$\frac{1}{2}$}}
edge from parent node[left,xshift=-5]{$\frac{1}{2}$}
}
child{node[solid node,label=above right:{$5$}]{}
child{node[hollow node,label=below:{$10$}]{} edge from parent node[left]{$\frac{1}{2}$}}
child{node[hollow node,label=below:{$0$}]{} edge from parent node[right]{$\frac{1}{2}$}}
edge from parent node[right,xshift=5]{$\frac{1}{2}$}
};
\end{tikzpicture}
\end{center}
\caption{Probability tree representation of two temporal lotteries with $T=2$}
\label{perulottery}
\end{figure}

I consider temporal lotteries that are
deterministic in the first period. Temporal lotteries $(D_t)_{t=0}^T$ are defined by $D_{T}:= C$ and recursively, $D_{t}:= C \times \Delta_b(D_{t+1}),$ for every $t=0,\ldots,T-1$. Likewise, simple temporal lotteries are defined by $D_{T,s}:= C$ and recursively $$D_{t,s}:= C \times \Delta_s(D_{t+1,s}),$$ for every $t=0,\ldots,T-1$. Simple temporal lotteries can be intuitively represented using a tree diagram, as illustrated in Figure \ref{perulottery}.

 I write $\left(c_0,\left(c_1, m\right)\right) \in D_0$ for a temporal lottery that consists of two periods of deterministic consumption, $c_0$ and $c_1$, followed by the lottery $m \in \Delta_b(D_{2})$. More generally, for any consumption vector $c^t=\left(c_0, \ldots, c_{t-1}\right) \in C^t$ and $m \in \Delta_b(D_{t})$, the temporal lottery $\left(c_0,\left(c_1,\left(c_2,\left(\ldots,\left(c_{t-1}, m\right)\right)\right)\right)\right)  \in D_0$ or $\left(c^t, m\right)$ for brevity is one that consists of $t$ periods of deterministic consumption followed by the lottery $m$.   Given two Polish spaces $X,Y$ and $m\in\Delta_b(X\times Y)$ I denote by $\operatorname{marg}_X m$ the marginal probability over $X$, i.e., $\operatorname{marg}_X m(A)=m(A\times Y)$ for every measurable set $A\subseteq X$.  
\begin{example}\label{ex:1}
Assume $T=2$. Let $d=(c_0,m)=\left(1,\frac{1}{2}\left(5,10\right)\oplus \frac{1}{2}\left(5,0\right)\right)$ and $d'=(c_0,m')=(1,5,(\frac{1}{2} 10  \oplus \frac{1}{2} 0))$.  Figure \ref{perulottery} provides a graphical representation of these two temporal lotteries.  We have $|\mathrm{supp}m|=2,|\mathrm{supp}m'|=1$ and   \[ M\left[\operatorname{marg}_{\Delta_s(D_{2,s})}  m'\right]=\begin{bmatrix}
    \frac{1}{2} & \frac{1}{2} \\
\end{bmatrix}
\text{ and  }
M\left[\operatorname{marg}_{\Delta_s(D_{2,s})}  m\right]=\begin{bmatrix}
    1 & 0 \\
    0 & 1\\
\end{bmatrix}.
\]

Moreover,  $\mu[\operatorname{marg}_{\Delta_s(D_{2,s})}  m']=[1]$ and $\mu[\operatorname{marg}_{\Delta_s(D_{2,s})}  m]=\left[ \frac{1}{2}\quad\frac{1}{2} \right]$.\demo
\end{example}

The preferences of a DM over temporal lotteries are given by a collection $(\succeq_{t})_{t=0}^T$ where each $\succeq_t$ is a weak order over $D_t$ and $\succ_t$ denotes the asymmetric part of $\succeq_t$. To ease notation, I denote by $\succeq:=(\succeq_t)_{t=0}^T$ the entire collection of preferences. 
\begin{definition}[Kreps-Porteus preferences]
Preferences $\succeq$ admit a \emph{Kreps-Porteus (KP) recursive representation} $(\phi, u, \beta)$  if each $\succeq_t$ is represented by $V_t: D_t \rightarrow \mathbb{R}$ such that $V_T(c) = u(c)$ for every $c \in C$ and recursively
$$
V_t(c, m)=u(c)+\beta \phi^{-1}\left(\mathbb{E}_m\phi\left(V_{t+1}\right)\right)\quad \text{for }t=0,\ldots,T-1,
$$ where $\beta \in (0,1]$ is a discount factor, $u : C \to \mathbb{R}$ is continuous and strictly increasing, with image either $u(C) = [0,\infty)$ or $u(C) = (0,\infty)$, and $\phi : u(C) \to \mathbb{R}$ is a continuous and strictly increasing function. Given a KP representation $(\varphi,u,\beta)$, the
\emph{risk-adjusted continuation utility} is given by
\[
\tilde{V}_t(c,m) := \mathbb{E}_m\phi\big(V_{t+1}\big),
\qquad t = 0,\ldots,T-1.
\]
\end{definition}
This representation of preferences separates risk aversion (captured by the function $\phi$) from the EIS (modeled by the utility function $u$). It is ordinally equivalent to more common formulations used in applications (see, e.g., \cite{werner2024ordinal}). The axiomatic foundations for this representation with bounded $u$ are well known (e.g., Proposition 4 in \citealt{sarver2018dynamic}); the unbounded case can be covered by results from \cite{bleichrodt2008koopmans}. The assumption of unboundedness is necessary  for the characterization provided later in Proposition \ref{peruchar}. The parameter $\beta$ is unique, whereas $u$ is cardinally unique, and $\phi$ is cardinally unique given $u$.\footnote{In applied literature, the function $\phi$ is often referred to as \textit{risk adjustment}; see, e.g., \cite{hansen2007intertemporal}. Moreover, $\beta = 1$ is permitted due to the finite horizon, which would not be the case under an infinite horizon.}  

Two notable cases are the \textbf{Epstein--Zin (EZ) preferences}, defined by
\begin{equation}\label{eq:EZ}
u(c)=\frac{c^\rho}{\rho},\quad \phi(x)=\frac{1}{\alpha}(\rho x)^{\frac{\alpha}{\rho}},\quad c\in C,\; x\in u(C),
\end{equation}
with parameters $0\neq \alpha<1$, $\rho\in (0,1)$, and $\alpha\leq \rho$; and the \textbf{Hansen--Sargent (HS) multiplier preferences}, given by
\begin{equation}
\phi(x)=-\exp\left(-\frac{x}{\theta}\right),\quad x\in u(C),
\end{equation}
with parameter $0<\theta<\infty$.\footnote{Under the present taxonomy, EZ preferences do not overlap with HS preferences, but they would if one allowed for $\rho=0$; see for example \cite{hansen2007intertemporal}, Example 2.3. In this case, when $C=(0,\infty)$ we have  $u(x)=\log(x)$ and $\phi(x)= -\exp{\left(\alpha {x}\right)}$ where $\alpha=-\frac{1}{\theta}$.}

I will typically  consider KP representations  that satisfy certain differentiability assumptions  to employ standard tools from the theory of risk aversion. Write $\phi\in \mathcal{C}^r$ if $\phi$   has  $r$ continuous derivatives. Given $\phi\in\mathcal{C}^2$, the  Arrow–Pratt index $A_{\phi}:\interior u(C)\rightarrow \mathbb{R}$ is given by
$$A_\phi(x)=-\frac{\phi''(x)}{\phi'(x)}\quad\text{ for every }x\in \interior u(C),$$
and the index of relative risk aversion  is  defined by
$R_\phi(x)=xA_{\phi}(x)$ for every $x\in \interior u(C)$.
 A function $\phi$ is decreasing absolute risk averse (DARA) if $A_\phi$ is non-increasing, it is increasing absolute risk averse (IARA) if its index $A_\phi$ is nondecreasing, and it is constant absolute risk averse (CARA) if it is both DARA and IARA. Increasing (IRRA), decreasing (DRRA), and constant (CRRA) relative risk averse functions are  defined analogously by replacing the index $A_\phi$ with $R_\phi$.
 \subsection{Preference for (non-instrumental) information}
To formally model the trade-off between intertemporal hedging and non-instrumental information, I reframe the theory of preferences for early resolution of uncertainty using the language of information economics. Temporal lotteries are partially ordered using a version of the Blackwell order, which allows them to be compared in terms of their (non-instrumental) informativeness. Beyond its theoretical appeal and generality,  this approach  permits building a formal link between correlation and information, which is central to the main results presented in the next section (see Proposition \ref{iidinfo}).

Similar to the Blackwell order, this ranking is based on the concept of \textit{garbling}. Consider $m, m^\prime \in \Delta_s(\Delta_s(X))$ such that $\cup_{\ell \in \mathrm{supp}m} \mathrm{supp}\ell = \cup_{\ell \in \mathrm{supp}m'} \mathrm{supp}\ell$, meaning they have the same support over terminal outcomes. Then $m^\prime$ is a garbling of $m$ if they can be associated with $(\mu[m], M[m])$ and $(\mu[m'], M[m'])$, where the columns of $M[m]$ and $M[m']$ represent the same outcomes, and there exists a stochastic matrix $G$ such that $M[m^\prime] = G M[m]$ and $\mu[m^\prime] G = \mu[m]$.
\begin{definition}[Temporal Blackwell]\label{def:infoblack}
Consider $d,d'\in D_{0,s}$ such that $d=\left(c^{t+1}, m\right),d'=\left(c^{t+1}, m'\right)$ for some $t\leq T-2$, $c^{t+1}\in C^{t+1}$, and  $m,m'\in \Delta_s(C\times \Delta_s(D_{t+2,s}))$  . Say that $d$ is more informative than $d'$, denoted $d\geq_B d'$, if  $\operatorname{marg}_{C} m'=\operatorname{marg}_{C} m$  and $\operatorname{marg}_{\Delta_s(D_{t+2,s})} m'$ is a garbling of $\operatorname{marg}_{\Delta_s(D_{t+2,s})} m$.
\end{definition}
In words, the expression $d\geq_B d'$ means that the two lotteries, $d$ and $d'$, have the same distribution of consumption in  period $t+1$. However, the actual realization of consumption in period $t+1$ provides more information about future values of consumption (from period $t+2$ onwards) for the lottery $d$ compared to the lottery $d'$. 

\begin{figure}
\begin{center}
\begin{tikzpicture}[scale=0.8,font=\footnotesize]
\tikzstyle{solid node}=[circle,draw,inner sep=1.5,fill=black]
\tikzstyle{hollow node}=[circle,draw,inner sep=1.5]
\tikzstyle{level 1}=[level distance=15mm,sibling distance=4cm]
\tikzstyle{level 2}=[level distance=15mm,sibling distance=1.5cm]
\tikzstyle{level 3}=[level distance=15mm,sibling distance=1cm]
\node(0)[solid node,label=above:{$0$}]{}
child{node[solid node,label=above left:{$1$}]{}
child{node[hollow node,label=below:{$1$}]{} edge from parent node[left]{$	1$}}
child{node[hollow node,label=below:{$0$}]{} edge from parent node[right]{$0$}}
edge from parent node[left,xshift=-5]{$\frac{1}{2}$}
}
child{node[solid node,label=above right:{$0$}]{}
child{node[hollow node,label=below:{$1$}]{} edge from parent node[left]{$0$}}
child{node[hollow node,label=below:{$0$}]{} edge from parent node[right]{$1$}}
edge from parent node[right,xshift=5]{$\frac{1}{2}$}
};
\end{tikzpicture}
\begin{tikzpicture}[scale=0.8,font=\footnotesize]
\tikzstyle{solid node}=[circle,draw,inner sep=1.5,fill=black]
\tikzstyle{hollow node}=[circle,draw,inner sep=1.5]
\tikzstyle{level 1}=[level distance=15mm,sibling distance=3.5cm]
\tikzstyle{level 2}=[level distance=15mm,sibling distance=1.5cm]
\tikzstyle{level 3}=[level distance=15mm,sibling distance=1cm]
\node(0)[solid node,label=above:{$0$}]{} 
child{node[solid node,label=above left:{$1$}]{} 
child{node[hollow node,label=below:{$1$}]{} edge from parent node[left]{$\frac{1}{2}$}}
child{node[hollow node,label=below:{$0$}]{} edge from parent node[right]{$\frac{1}{2}$}}
edge from parent node[left,xshift=-5]{$\frac{1}{2}$}
}
child{node[solid node,label=above right:{$0$}]{}
child{node[hollow node,label=below:{$1$}]{} edge from parent node[left]{$\frac{1}{2}$}}
child{node[hollow node,label=below:{$0$}]{} edge from parent node[right]{$\frac{1}{2}$}}
edge from parent node[right,xshift=5]{$\frac{1}{2}$}
};
\end{tikzpicture}
\end{center}
\caption{Probability tree representation of a temporal lottery}
\label{iidtemporal}
\end{figure}
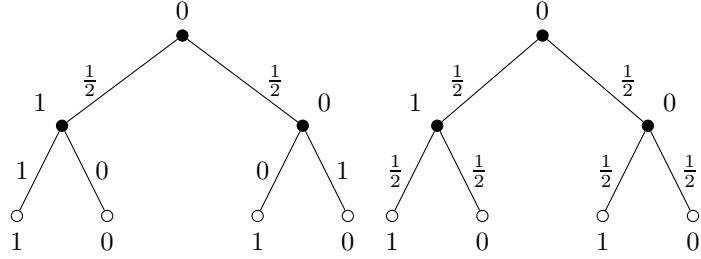

Observe that $\geq_B$ is a partial order just like the standard Blackwell order. A full characterization of this order is left for future research. A natural starting point would be to adopt the methods from \cite{kihlstrom1984bayesian}. The following examples illustrate this notion of comparative information.
\begin{example}[Example \ref{ex:1} continued]\label{example1}
Recall that here we have $d=\left(1,\frac{1}{2}\left(5,10\right)\oplus \frac{1}{2}\left(5,0\right)\right)$ and $d'=(1,5,(\frac{1}{2} 10  \oplus \frac{1}{2} 0))$.   If we let $G=\begin{bmatrix}
    \frac{1}{2} & \frac{1}{2} 
\end{bmatrix}$
\[ M\left[\operatorname{marg}_{\Delta_s(D_{2,s})}  m'\right]=\begin{bmatrix}
    \frac{1}{2} & \frac{1}{2} \\
\end{bmatrix}=\begin{bmatrix}
    \frac{1}{2} & \frac{1}{2} \\
\end{bmatrix}\begin{bmatrix}
    1 & 0 \\
    0 & 1\\
\end{bmatrix}=
G M\left[\operatorname{marg}_{\Delta_s(D_{2,s})}  m\right],\]
and
\[\mu[\operatorname{marg}_{\Delta_s(D_{2,s})}  m']G=\begin{bmatrix}
   1 
\end{bmatrix}\begin{bmatrix}
    \frac{1}{2} & \frac{1}{2} 
\end{bmatrix}=\begin{bmatrix}
    \frac{1}{2} & \frac{1}{2} 
\end{bmatrix}=\mu[\operatorname{marg}_{\Delta_s(D_{2,s})}  m].
\]

Furthermore, $\operatorname{marg}_{C} m'=\operatorname{marg}_{C} m=\delta_{5}$,  so that $d\geq_B d'$. In words, the terminal value of consumption is fully revealed by a coin toss at $t=1$ for $d$ but only  revealed  at $t=2$   for $d'$.\demo
\end{example}
\begin{example}\label{example2}
Again assume $T=2$. Consider $d=(1,m),d'=(1,m')$ given by $d=\left(1,\frac{1}{2}\left(1,1\right)\oplus \frac{1}{2}\left(0,0\right)\right)$ and $d'=\left(1,\frac{1}{2}\left(1,\left(\frac{1}{2} 1  \oplus \frac{1}{2} 0\right)\right)\oplus \frac{1}{2}\left(0,\left(\frac{1}{2} 1  \oplus \frac{1}{2} 0\right)\right)\right)$.   Figure \ref{iidtemporal} provides a graphical representation of these two temporal lotteries.  We have   \[ M\left[\operatorname{marg}_{\Delta_s(D_{2,s})}  m'\right]=\begin{bmatrix}
    \frac{1}{2} & \frac{1}{2} \\
    \frac{1}{2} & \frac{1}{2}\\
\end{bmatrix}=
\begin{bmatrix}
    \frac{1}{2} & \frac{1}{2} \\
    \frac{1}{2} & \frac{1}{2} \\
\end{bmatrix}
\begin{bmatrix}
    1 & 0 \\
    0 & 1\\
\end{bmatrix}=
\begin{bmatrix}
    \frac{1}{2} & \frac{1}{2} \\
    \frac{1}{2} & \frac{1}{2} \\
\end{bmatrix}
M\left[\operatorname{marg}_{\Delta_s(D_{2,s})}  m\right],
\]which implies that $m'$ is a garbling of $m$. Furthermore,  $\operatorname{marg}_{C} m'=\operatorname{marg}_{C} m$, so that $d\geq_B d'$. In words, $d'$ is an ``iid'' temporal lottery while $d$ is perfectly correlated.\demo
\end{example}
I follow \cite{bommier2017monotone} to describe an agent who exhibits a preference for early resolution of uncertainty  (see their Definition 2 and also \citealt{strzalecki2013temporal}).
\begin{definition}[PERU]\label{def:peru}
Preferences $\succeq$ exhibit a preference for early resolution of uncertainty (PERU)  if  for every $n>0$,  $c_0,c_1\in C$, $(c_{2i},m_i)_{i=1}^n\in D_{2,s}$,   and probability vector $\left(\pi_i\right)_{1 \leq i \leq n}$
$$d_{KP}^{early}:=\left(c_0, \bigoplus_{i=1}^n \pi_i\left(c_1, c_{2 i}, m_i\right)\right) \succeq_0 \left(c_0, c_1, \bigoplus_{i=1}^n \pi_i\left(c_{2 i}, m_i\right)\right):= d_{KP}^{late}$$
\end{definition}

As I show in the Appendix (see Lemma \ref{lemma:tedeum}), $d_{KP}^{early}\geq_B d_{KP}^{late}$, which explains why preferring   $d_{KP}^{early}$ to $d_{KP}^{late}$ can equivalently be understood as a preference for non-instrumental information.    The next result characterizes PERU. 
\begin{proposition}\label{peruchar}
Assume $\succeq$ admit a KP representation $(\phi,u,\beta)$ with $\phi\in \mathcal{C}^2$. Then preferences $\succeq$ exhibit PERU 
if and only if
\begin{equation}\label{convexity}
-\beta \frac{\phi^{\prime \prime}(\beta x+y)}{\phi^{\prime}(\beta x+y)} \leq- \frac{\phi^{\prime \prime}(x)}{\phi^{\prime}(x)},
\end{equation}
for every $x,y\in \interior u(C)$.\footnote{Condition \eqref{convexity} is due to \citeauthor{strzalecki2013temporal} \citeyearpar{strzalecki2013temporal}; see p. 1051.}
\end{proposition}
\begin{proof}
See the Appendix.
\end{proof}
Observe that the quantity defined by 
\begin{equation}\label{equation:er} 
ER_\phi(x,y)=-\frac{\phi^{\prime \prime}(x)}{\phi^{\prime}(x)}+\beta \frac{\phi^{\prime \prime}(\beta x+y)}{\phi^{\prime}(\beta x+y)}\quad\text{for every }x,y\in \interior u(C),
\end{equation}
 can be considered as a local measure of strength of preference for non-instrumental information. In Appendix \ref{subsec:measuringperu}, I introduce the notion of \textit{early resolution premium}, and show that it is a function of the weighted average of different values of $ER_\phi$. To illustrate, when $\beta=1$ and $\phi(x)= -\exp \left(-\frac{x}{\theta}\right)$ we obtain that 
   
 $$ER_\phi(x,y)=\frac{1}{\theta}-\frac{1}{\theta}=0,
   $$ which implies indifference to non-instrumental information. The same applies if $\phi$ is the identity.

 Here I focus on  risk attitudes that exhibit a preference for information regardless of the level of impatience or intertemporal substitution. 
\begin{definition}[UPI] Say that $\phi$ satisfies a uniform preference for information (UPI) if  every preference relation $
\succeq$ with a KP representation $(\phi,u,\beta)$ exhibit PERU.
\end{definition} 
The next simple result provides a connection between classical risk attitudes and  preference for information.
\begin{proposition}\label{prop:upidara}
If $\phi\in \mathcal{C}^2$ satisfies UPI  then it also satisfies DARA. 
\end{proposition}
\begin{proof}
Immediate from  \eqref{convexity}.
\end{proof}

\section{Main results: correlation aversion}\label{attcorr}
 I introduce a general notion of an increase in positive correlation between consumption at two distinct periods. I then characterize recursive preferences that are averse to correlation. For ease of exposition, I consider the  case in which there are two risky periods, i.e., $T=2$. The Supplemental Appendix (see Section  \ref{infinite}) extends the results  to an infinite horizon $T=\infty$. 

I introduce a class of temporal lotteries that can be defined by (i) the distribution of consumption at time $t=1$ and (ii) the conditional distribution of consumption at time $t=2$ given consumption in the previous period. The advantage is that lotteries within this class   can be ordered based on their correlation.

Let $$M_s^*:=\left\{m\in\Delta_s(C \times \Delta_{s}(C)):(c,\mu),(c,\mu')\in\operatorname{supp}m\implies  \mu= \mu'\right\}.$$
Every such $m\in M_s^*$ can be (uniquely) associated  with $m_1\in\Delta_s(C)$ and $m_2(\cdot|\cdot)\in\Delta_s(C)^{\mathrm{supp}m_1}$, defined by $m_1=\operatorname{marg}_C m,$ and $$m_2(\cdot|c)=\mu(\cdot),$$
where $\mu$ is the unique element of $\Delta_s(C)$ such that $(c,\mu)\in \operatorname{supp} m$.  Conversely, given $m_1\in\Delta_s(C)$ and $m_2(\cdot|\cdot)\in\Delta_s(C)^{\mathrm{supp}m_1}$, we can uniquely define $m\in M^*_s$ by $$m(c,m_2(\cdot|c)):=m_1(c)\quad\text{for every } c\in \mathrm{supp}m_1.$$  In words, $m_1$ describes the distribution of time $1$ consumption while $m_2(\cdot|c)$ is the conditional distribution of consumption at the final time period given a realization of $t=1$ consumption. The set $D_{0,s}^{*}:=\left\{(c,m)\in D_{0,s}:m\in M_s^*\right\}$  is the set of temporal lotteries that can be described in terms of a pair $(m_1,m_2)$.

The structure of these lotteries can be used to introduce the following notion of increasing correlation.
\begin{definition}[IECIT] Consider $d=(c_0,m),d'=(c_0,m')\in D_{0,s}^*$. Say that $d$ differs from $d'$ by an \textbf{intertemporal elementary correlation increasing transformation (IECIT)}   if  and only if $m_1=m_1'$ and there exist  $\varepsilon\geq 0$ and a pair $(c,c')$ such that $c\neq c', m_1(c),m_1(c')\neq 0$ and $$m_2(c|c)=m_2'(c|c)+ \frac{\varepsilon}{m_1'(c)},$$
 $$m_2(c'|c)=m_2'(c'|c)- \frac{\varepsilon}{m_1'(c)},$$
   $$m_2(c'|c')=m_2'(c'|c')+ \frac{\varepsilon}{m_1'(c')},$$
 $$m_2(c|c')=m_2'(c|c')- \frac{\varepsilon}{m_1'(c')},$$
   and $m_2=m_2'$ otherwise. 
\end{definition}
In simpler terms, these transformations increase the probability that if consumption at $t=1$ is either $c$ or $c'$ it will remain the same at $t=2$ and concurrently decrease the probability that consumption will shift to a different level.   The following two examples serve as an illustration of this concept.
\begin{example}[Example \ref{example2} continued]\label{example3} In this case we have $m_1=m_1'$, $m_2(1|1)=1=m_2'(1|1)+\frac{1}{1/2}\frac{1}{4}=\frac{1}{2}+\frac{1}{2}$, $m_2(0|1)=0=m_2'(0|1)-\frac{1}{1/2}\frac{1}{4}=\frac{1}{2}-\frac{1}{2}$, $m_2(1|0)=0=m_2'(1|0)-\frac{1}{1/2}\frac{1}{4}=\frac{1}{2}-\frac{1}{2}$ and $m_2(0|0)=1=m_2'(0|0)+\frac{1}{1/2}\frac{1}{4}=\frac{1}{2}+\frac{1}{2}$. It follows that  $d$ differs from $d'$ by an IECIT with $\varepsilon=\frac{1}{4}$. Therefore, the perfectly correlated temporal lottery $d$ can be obtained from the ``iid'' lottery $d'$ by means of an IECIT. In this case, an IECIT also increases the informativeness of a temporal lottery.\demo
\end{example}

The concept of an IECIT is an application of \citeauthor{epstein1980increasing}'s (\citeyear{epstein1980increasing}) idea of generalized increasing correlation, applied in a dynamic setting. With the notion of an IECIT, it is possible to establish an ordering $\geq_C$ that can be used to rank temporal lotteries based on their persistence.
\begin{definition}[Correlation order]
Given $d,d^{\prime} \in D_{0, s}^*$ say that $d$ is more correlated than $d'$, denoted $d\geq_C d'$, if   $d$ differs from $d'$  by a finite amount of IECITs.
\end{definition}
Observe that $\geq_C$ is transitive and thus a partial order. The following result establishes a formal connection between IECITs and non-instrumental information by showing that increasing the correlation of ``iid'' temporal lotteries makes them more informative. To this end, define the iid temporal lottery for each $\ell \in \Delta_s(C)$  by $d^{iid}(\ell) = (c,m)$ where $m_1(\cdot)=m_2(\cdot|c)=\ell(\cdot)$ for every $c\in C$. In words, the distribution of consumption at each period is described by $\ell$. 
\begin{proposition}\label{iidinfo}
Consider $\ell\in\Delta_s(C)$ and $d,d'\in D_{0,s}^*$.  Then it holds that $$d\geq_C d'\geq_C d^{iid}(\ell)\implies d\geq_B d'\geq_B d^{iid}(\ell).$$
\end{proposition}
\begin{proof}See the Appendix.
\end{proof}
This proposition establishes formally the main trade-off described in the introduction: increasing persistence in consumption risks to an iid lottery provides more information about future consumption. We can define correlation aversion as aversion towards increasing correlation to an  iid temporal lottery.

\begin{definition}[Correlation aversion]\label{correlationaversionaxiom}
Preferences $\succeq$ exhibit correlation aversion if  for every $d,d'\in D_{0,s}^*$ and $\ell\in\Delta_{s}(C)$ $$d\geq_C d'\geq_C d^{iid}(\ell)\implies d^{iid}(\ell)\succeq_0 d'\succeq_0 d.$$
\end{definition}

The next result characterizes correlation-averse preferences in terms of risk attitudes, under the assumption of UPI, i.e., when there is a  trade-off between intertemporal hedging and non-instrumental information.
 \begin{theorem}\label{theo1}
Consider $\phi\in \mathcal{C}^3$ that  satisfies UPI. Then every preference relation $\succeq$ with KP representation $(\phi,u,\beta)$ exhibit correlation aversion if and only if $\phi$ is concave and satisfies IRRA.
\end{theorem}
 \begin{proof}
See the Appendix.
 \end{proof}
Thus, when $\phi$ satisfies UPI, every  KP preference with representation $(\phi,u,\beta)$ is correlation-averse exactly when it is risk averse (i.e., $\phi$ is concave) and exhibits increasing relative risk aversion (IRRA).  Observe  that IRRA is one of the most important
classes of utility functions  (e.g., see \cite{arrow1971essays}, p. 96), and notably includes the Epstein–Zin   and Hansen–Sargent  preferences. Moreover, empirical findings support DARA
and IRRA (\cite{wakker2010prospect}, p. 83). This result further implies that indifference to correlation occurs only under an affine adjustment factor, i.e., any positive affine transformation of  $\phi(x)=x$.\footnote{Note that the ``boundary'' case of indifference to correlation  occurs  when the relative risk aversion function $R_\phi$ is constant and equal to zero. Indeed, as a consequence of the proof of Theorem \ref{theo1}, whenever $R_\phi(x)>0$ for some $x$, one can construct an iid lottery that is strictly better than a lottery that differs from it by an IECIT.}

\vspace{0.15cm}

\noindent \textbf{A bound on preference for information}. A central implication of Theorem \ref{theo1} is that IRRA constrains preference for information. To illustrate, assume that $\phi$ exhibits HARA: suppose that  for $x>1$ $$
\phi(x)=\frac{1-\gamma}{\gamma}\left(\frac{x}{1-\gamma}+b\right)^\gamma, \quad \text{with } 0 \neq \gamma<1 \text{ and } b\in \left[\frac{1}{\gamma-1},\infty\right).
$$
Then, given $x,y>1$, it is easy to check that when $\beta=1$ the   local measure of strength of preference for non-instrumental information $$-\frac{\phi^{\prime \prime}(x)}{\phi^{\prime}(x)}+ \frac{\phi^{\prime \prime}( x+y)}{\phi^{\prime}( x+y)},$$ 
is  decreasing in the parameter $b$. In words, the smaller the parameter 
$b$, the greater the preference for  information. In this case, we have $$R_{\phi}^\prime(x)=\frac{(\gamma-1)^2 b}{\left(b(1-\gamma)+x\right)^2},$$ so that IRRA implies  $b\geq 0$, thus excluding the case $b\in  \left[\frac{1}{\gamma-1},0\right)$, where the DM values information more. Therefore, IRRA  limits preference for information.

More generally, note that  IRRA means  that $R_{\phi}$ is non-decreasing. Therefore, when $R_{\phi}$ is differentiable  we have:$$R_{\phi}'(x)\geq 0\implies A_{\phi}'(x)\geq -\frac{A_{\phi}(x)}{x},$$
for every $x\neq 0$.  Under DARA it holds $A_{\phi}'\leq 0$, so that we obtain  
$$A_{\phi}'(x)\in \left[-\frac{A_{\phi}(x)}{x},0\right].$$
This means that IRRA limits the reduction of absolute risk aversion for a given increase in utility. Therefore, when $\beta$ is close to unity, IRRA effectively imposes an upper bound on $ER_\phi(x,y)$ since   
\begin{equation}\label{eq:boundperu}
ER_\phi(x,y)=-\frac{\phi^{\prime \prime}(x)}{\phi^{\prime}(x)}+\beta \frac{\phi^{\prime \prime}(\beta x+y)}{\phi^{\prime}(\beta x+y)}\approx -A_{\phi}'(x)y\leq \frac{A_{\phi}(x)y}{x}.
\end{equation}

\vspace{0.15cm}
\noindent
\textbf{The persistence premium}. To measure a DM's level of correlation aversion, I introduce the notion of the persistence premium. The persistence premium quantifies how much a DM is willing to pay to eliminate all persistence from consumption. As a byproduct of Theorem \ref{theo1}, I derive an approximation of the persistence premium in the spirit of \cite{pratt1976risk} and \cite{bommier2007risk}, which connects the premium with preference for information.

Formally,  given $c_0>0,x>y> 0,$ let

\[
d^{corr}(\varepsilon) = \scalebox{0.9}{$
\left(c_0, \frac{1}{2}\left( x, \left( \left( \frac{1}{2} + \frac{\varepsilon}{2} \right)x \oplus \left( \frac{1}{2} - \frac{\varepsilon}{2} \right)y \right) \right) \oplus \frac{1}{2} \left( y, \left( \left( \frac{1}{2} - \frac{\varepsilon}{2} \right)x \oplus \left( \frac{1}{2} + \frac{\varepsilon}{2} \right)y \right) \right) \right)$},
\]
  
  and    
\[
d^{iid}(\pi) = \left(c_0, \scalebox{0.85}{$\frac{1}{2} \left( x(1-\pi), \left( \frac{1}{2} x(1-\pi) \oplus  
\frac{1}{2} y(1-\pi) \right) \right) \oplus \frac{1}{2} \left( y(1-\pi), 
\left( \frac{1}{2} x(1-\pi) \oplus \frac{1}{2} y(1-\pi) \right) \right)$} \right).
\]

The correlated lottery $d^{corr}(\varepsilon)$ is a generalization to arbitrary consumption levels  of the perfectly correlated lottery in Example \ref{example2}. In this case, the level of correlation depends on the parameter $\varepsilon\in [0,1]$, where at $\varepsilon=0$ one has no correlation and perfect correlation at $\varepsilon=1$. In contrast, the  lottery $d^{iid}(\pi)$ is iid   but the payoffs are discounted by a factor of $(1-\pi)\in [0,1]$.

  Consider  preferences $\succeq$ with   KP representation $(\phi,u,\beta)$, where  $\phi\in \mathcal{C}^3$  is concave,  and satisfies both IRRA  and  UPI. For simplicity, assume that $u(x)=x$. By Theorem  \ref{theo1}, $V_0\left(d^{iid}(0)\right)\geq V_0\left(d^{corr}\left(\varepsilon\right)\right)$ for every $\varepsilon\in [0,1]$, and as we increase $\pi$, $V_0\left(d^{iid}\left(\pi\right)\right)$ strictly decreases. In particular, $V_0\left(d^{iid}(1)\right)< V_0\left(d^{corr}\left(\varepsilon\right)\right)$ for every $\varepsilon\in [0,1]$.  Therefore,  we can denote by $\pi(\varepsilon)\in [0,1]$  the  unique solution to the equation 
  \begin{equation}\label{eq:corrpremium}V_0\left(d^{corr}\left(\varepsilon\right)\right)=V_0\left(d^{iid}\left(\pi\left(\varepsilon\right)\right)\right).
  \end{equation}
The persistence premium $\pi(\varepsilon)$ therefore quantifies how much of consumption one is willing to relinquish to have all persistence removed from consumption. The next result provides an approximation of $\pi(\varepsilon)$ near $\varepsilon=1$.
 \begin{corollary}\label{premiumcorollary}
Fix $(c_0,x,y,\beta,\phi)$. There exist constants $k_1,k_2,k_3>0$, depending only on
$(c_0,x,y,\beta,\phi)$, such that 
\begin{small}
\begin{align}
\pi(\varepsilon) \approx & \, k_1\big(\tilde{V}_0(d^{iid}(0)) - \tilde{V}_0(d^{corr}(1))\big) \nonumber \\
& + k_2 (\varepsilon - 1) \int_{y}^{x} \frac{\phi'(z(1+\beta))}{\phi'(z)}
\frac{\big\{R_{\phi}(z(1+\beta)) - R_{\phi}(z)\big\}}{z} \, dz   \label{eq:appropremium} \\
& - k_3(\varepsilon - 1)^2 \left\{\frac{\phi'(x(1+\beta))}{\phi'(x)} ER_{\phi}(x,x)
+ \frac{\phi'(y(1+\beta))}{\phi'(y)} ER_{\phi}(y,y)\right\} \nonumber .
\end{align}
\end{small}
\end{corollary}
 \begin{proof}
 See the Appendix.
 \end{proof}
The first term of this approximation shows that the persistence premium
positively depends on the difference between the value of the iid and the
perfectly correlated lottery, independently of the level of correlation
$\varepsilon$. In parametric formulations, such as EZ or HS preferences,
the term $\tilde{V}_0(d^{iid}(0)) - \tilde{V}_0(d^{corr}(1))$ increases
with the parameter of risk aversion.

The second and third terms capture, respectively, the local ``speed'' and
``acceleration'' of the persistence premium as the level of correlation
changes around $\varepsilon=1$. Since $\phi$ satisfies IRRA,
$R_{\phi}(z(1+\beta)) - R_{\phi}(z)$ is non-negative, and since additionally
$\phi'>0$ (and $z>0$) the integral in the second term is positive. Hence,
approximately, the premium is increasing in the level of
persistence. In particular, the larger is the increase of $R_{\phi}$ with
consumption, the faster the premium grows with persistence. By contrast, the
third term depends on an average of the measure of preference for early
resolution of uncertainty, $ER_\phi(x,x)$ and $ER_\phi(y,y)$. A stronger
preference for early resolution (higher $ER_\phi$) makes the premium grow at
a more rapidly decreasing pace. Hence this local approximation reflects the
trade-off between hedging and information.

To illustrate this approximation in a practical case, quick calculations reveal that in the EZ model where $\phi(x)=\frac{1}{\alpha}(\rho x)^{\frac{\alpha}{\rho}}$, we
obtain $$ER_\phi(x, y)=\frac{y\left(1-\frac{\alpha}{\rho}\right)}{x(\beta x+y)},$$
and, for some $\tilde{a}, \tilde{b}, \tilde{c} > 0$,
\[
\begin{aligned}
\pi(\varepsilon) \approx\;
  \tilde{a}
  + \tilde{b}\,\varepsilon \Bigl(1 - \frac{\alpha}{\rho}\Bigr)
      \Bigl(\frac{1}{x} + \frac{1}{y}\Bigr)
  \;-\;
  \tilde{c}\,(\varepsilon^2 - 1)\,ER_{\phi}(x,y)
  \;+\; \text{higher-order terms in }(\varepsilon-1).
\end{aligned}
\]
This expression is the EZ specialization of the approximation introduced
earlier. It illustrates how the trade-off depends on the preference
parameters: higher risk aversion causes the premium to rise more rapidly as
$\varepsilon$ increases, while higher EIS slows it when $\alpha < 0$ but
speeds it up when $\alpha > 0$. At the same time, a higher $ER_{\phi}$
moderates the increase. This moderating effect itself depends on both risk
aversion and EIS---since $ER_{\phi}$ depends on both $\alpha$ and $\rho$---
reflecting the fact that correlation aversion and PERU are not fully
distinguishable in the EZ model. I examine the implications of this result
for asset pricing in Section~\ref{impliforassetpricing}.

\vspace{0.15cm}
\noindent \textbf{Correlation aversion and model misspecification}. A further examination of IRRA reveals a tight connection between correlation aversion and fear of model misspecification.  Consider the following condition which strengthens  IRRA by requiring that the index of relative risk aversion increases sufficiently rapidly.\begin{definition}[SCA]
Say that $\phi\in\mathcal{C}^4$ satisfies strong correlation aversion (SCA) if  it  is concave, it satisfies IRRA, and $R_{\phi}''(x)\geq 0$ for every $x\in (0,\infty)$.
\end{definition}
 Thus, SCA requires  not only that the index of relative risk aversion $R_{\phi}$ is increasing, but also that it increases at a sufficiently fast pace.  Observe that both EZ and HS preferences satisfy this condition since in both cases $R_{\phi}''=0$. Proposition \ref{suppapp:SCA} in the Supplemental Appendix provides an axiomatic foundation for SCA.

To talk about model misspecification, one needs a notion of distance between probabilistic models. A function $I : X \times X \rightarrow [0, \infty]$ is called a \textit{statistical distance} \citep{liese1987convex} if $I(\cdot \| x)$ is convex and lower semicontinuous, and $I(x \| x) = 0$ for all $x \in X$. 
\begin{theorem}\label{theo2}
Assume that $\succeq$  admits a KP representation $(\phi,u,\beta)$ with $\phi\in\mathcal{C}^4$ that satisfies UPI. If $\phi$ satisfies SCA, then $\succeq$ admits the recursive representation $(V_t)_{t=0}^2$ which satisfies $V_2(c)=u(c)$ and for every $(c,m)\in D_{t,s}$
$$
V_t(c, m)=u(c)+\beta \min _{\ell\in\Delta_b(D_{t+1})}\left\{\mathbb{E}_{\ell} V_{t+1}+I_{(\phi,u,\beta)}^t(\ell|| m)\right\}\quad \text{for }t=0,1,
$$
 where each $I_{(\phi,u,\beta)}^t(\cdot, \cdot):\Delta_b(D_{t+1})\times \Delta_b(D_{t+1})\rightarrow [0,\infty]$  is a statistical distance.
\end{theorem}
\begin{proof}
See the Appendix.
\end{proof}

This result formalizes the connection between robustness to model misspecification and correlation aversion. By introducing a mild strengthening of risk attitudes related to correlation aversion, recursive preferences naturally reflect a fear of model misspecification. The extent to which SCA is necessary for this representation is discussed in the proof.

The intuition behind this representation is that the decision maker is concerned about the potential misspecification of the distribution of future consumption. As a result, alternative distributions are evaluated based on their distance from \(m\), as measured by the statistical distance \(I_{\phi, u, \beta}^t\) (the general formulation of these statistical distances is  discussed in the Appendix). Therefore, $I_{\phi, u, \beta}^t(\ell \| m)$ quantifies the ``cost'' of considering alternative distribution $\ell$ at time $t$. The dependence on \(t\) arises from the finite horizon, while the Supplemental Appendix establishes uniqueness when $T=\infty$.

As with multiplier preferences \citep{strzalecki2011axiomatic}, these statistical distances represent the fear of model misspecification---lower values imply lower costs for alternative distributions. However, unlike the framework in \cite{strzalecki2011axiomatic}, here the cost evaluates discrepancies between temporal lotteries rather than subjective beliefs.

For HS preferences, where $\phi(x) = -e^{-\frac{x}{\theta}}$, the cost function corresponds to Relative Entropy \citep{strzalecki2011axiomatic}:
\[
I_{\phi,u,\beta}^t(\ell \| m) = \theta \mathbb{E}_{m}\left[\frac{d \ell}{d m} \log \frac{d \ell}{d m}\right], \quad \text{if } \ell \ll m, \quad \text{else } \infty.
\]

Unlike HS preferences, the cost function for EZ preferences depends additionally on continuation utility. Moreover, \cite{meyer2019generalized} shows the cost function in this case can be expressed via the Tsallis entropy (see also equation \ref{eq:renyicost} for an equivalent characterization in terms of Rényi divergence). Theorem \ref{theo2} extends this  model misspecification interpretation to all recursive  preferences satisfying SCA. 



\section{Applications}
\subsection{Asset pricing}
\label{impliforassetpricing}
A key finding in macro-finance is that stocks are procyclical---they generate high returns during economic expansions but tend to crash in recessions (see, for example, \citealt{cochrane2008financial}). As a result, stocks are poor hedging instruments compared to bonds, which offer a relatively stable yield regardless of the business cycle. However, this limited hedging capability alone does not justify the substantial excess returns that investors require for holding equities. This inconsistency is known as the equity premium puzzle.

When economic growth is persistent, correlation aversion makes equity an even worse hedging instrument: stock prices rise with favorable long-term economic prospects and fall sharply when those prospects deteriorate. Consequently, Theorem \ref{theo1} and Corollary \ref{premiumcorollary} imply that investors who are correlation-averse will demand an even higher premium for bearing equity risk relative to bonds. In general, the relative attractiveness of bonds versus stocks depends on the strength of the preference for early resolution compared to risk aversion. Investors who are risk averse but place little value on early information will tend to favor bonds, whereas those who value early resolution strongly relative to their risk aversion will prefer stocks, since stocks convey news about long-run consumption growth.

This observation carries important macro‐financial implications. In particular, the long-run risk  model  of \cite{bansal2004risks}  introduces a persistent component in  consumption growth. If the  representative investor  has \citeauthor{epstein1989substitution}'s preferences, this model is able to match the observed  equity premium. In particular, this model relies on a  consumption process (case I) that satisfies for  $t=0,\ldots$
\begin{align}\label{LRR}
\begin{split}
\log \left(\frac{c_{t+1}}{c_{t}}\right) =m+x_{t+1}+\sigma \epsilon_{c, t+1}, \\
x_{t+1} =a x_{t}+\varphi \sigma \epsilon_{x, t+1}, \\
\epsilon_{c, t+1},  \epsilon_{x, t+1} \sim \text { iid } N(0,1),
\end{split}\end{align} where $d_t:=\log\left(c_{t+1}/c_{t}\right)$ denotes consumption growth, $m,\sigma,\phi>0$, and $a\in [0,1)$ is the persistence parameter. The representative investor has EZ preferences with discount factor $\beta\in (0,1)$ so that the utilities $U_t$   satisfy the recursion  
$$U_t=u^{-1}\left((1-\beta) u(c_t)+\beta u\left(\left[\mathbb{E}_t\left(U_{t+1}^\alpha\right)\right]^{1 / \alpha}\right)\right),$$
where $u(x)=x^\rho$ for $\rho\in [0,1)$, under the convention that for $\rho=0$, $u(x)=\log(x)$.  This formulation of EZ preferences  is ordinally equivalent to the one  in \eqref{eq:EZ} (e.g., see \citealt{werner2024ordinal}).

Such a model faces the trade-off discussed previously.    An investor with recursive preferences values both the early resolution of uncertainty and intertemporal hedging. Theorem \ref{theo1} and Corollary \ref{premiumcorollary} suggest that an investor with EZ preferences is worse off as the  persistence $a$ increases, but due to preference for information, at a decreasing rate. It follows that the persistent component of consumption inflates the equity premium because of correlation aversion, \textit{despite} preference for non-instrumental information. Preferences for early resolution of uncertainty still play a role in other contexts such as the macroeconomic announcement premium; see for example \cite{ai2018risk}.\footnote{Note that in the long-run risk model, there are other features of preferences at play, unrelated to those discussed here. For instance, in this model it is crucial that  $\mathrm{EIS}>1$, as this ensures that the substitution effect outweighs the income effect.}

\vspace{0.15cm}


\noindent
\textbf{The persistence premium under long-run risk}. 
 \citeauthor*{epstein2014much} (\citeyear{epstein2014much}) suggest that the long-run risk model entails implausibly high levels of preferences for early resolution of uncertainty. They introduce the concept of a ``timing premium'' to reflect  preferences for early resolution of uncertainty,  which, when \( \rho = 0 \), is given by:
$$
1-\exp \left\{\frac{\alpha}{2} \frac{\beta^2 \sigma^2}{1-\beta^2}\left(1+\frac{\varphi \beta^2}{(1-\beta a)^2}\right)\right\}.
$$ 
  Under the standard parameters of the model from the literature, they note that the resulting timing premium seems excessively high compared to introspective assessments.

In light of my analysis of correlation aversion, I ask a different question: ``What fraction of your wealth would you give up to remove all persistence in consumption?'' Formally, here  the persistence premium is defined as $$\pi:=1-\frac{U_0(d^{corr})}{U_0(d^{iid})},$$
where  $d^{iid}$ and  $d^{corr}$ are the stochastic processes of consumption in \eqref{LRR}   with $a=0$ (no persistence) and  $a=0.9790$, respectively. Given that EZ utility is positively homogeneous, the persistence premium, as in the previous section, quantifies the proportion of consumption an investor would be willing to forgo to eliminate all persistence from consumption.

 When  $\rho=0$ the persistence premium is given by (see  Appendix \ref{app:assetpricing}) $$\pi=1-\exp\left\{{\frac{\beta}{1-\beta a} x_{0}-\beta x_0+\frac{\alpha}{2} \frac{\beta \sigma^{2}}{1-\beta}\left(\frac{\varphi^{2} \beta^{2}}{(1-\beta a)^{2}}-\varphi^{2} \beta^{2}\right)}\right\},$$ 
\begin{table}
\begin{center}
 \caption{Parameters of the LRR model (see \citealt{epstein2014much})}
\begin{tabular}{|p{1cm}|p{1cm}|p{1cm}|p{1cm}|p{1cm}|p{1cm}|p{1cm}|p{1cm}|}
 \hline
   $\sigma$ & $\varphi$ &$a$ &$\beta$& $1-\alpha$ & $\rho$ & $x_0$ & $\pi$\\
 \hline
$0.0078$& 0.044 &   $0$ &  $0.998$ &   $7.5$&   $0$& $0$ & $0$ \\
\hline
$0.0078$& 0.044 &   $0.9790$ &  $0.998$ &   $7.5$&   $0$& $0$& $30\%$ \\
\hline
$0.0078$& 0.044 &   $0.9790$ &  $0.998$ &   $10$&   $0$& $0$& $40\%$ \\
 \hline
\end{tabular}
  \label{LRRparameters}
\end{center}
\end{table}
Thus,  $\pi$ increases with the degree of persistence $a$, and grows more rapidly with higher risk aversion $1 - \alpha$, as well as with the volatility parameters $\sigma$ and $\varphi$. Moreover, the timing premium also depends on these parameters in a similar fashion.

 This pattern mirrors  that of the approximate persistence premium discussed in the preceding section, based on the approximation in equation~\eqref{eq:appropremium}. In this case, when $\phi(x)=-e^{\alpha x}$, for some constants $\tilde{a}, \tilde{b} > 0$, and for values of $\beta$ close to unity, we have that for $x > y$:
\[
\pi(\varepsilon) \approx \tilde{a} - \tilde{b} \varepsilon \,\bigl(e^{\alpha x} - e^{\alpha y}\bigr).
\]
Hence, this premium also increases with the persistence parameter $\varepsilon \in [0,1]$. Moreover, for sufficiently small $\alpha$, the rate of increase is amplified by higher risk aversion $1 - \alpha$ and by greater consumption volatility, which is driven by the difference between $x$ and $y$.

Table \ref{LRRparameters} summarizes the parameters of the model. In particular, with a risk aversion level of \( 1 - \alpha = 7.5 \), I obtain \( \pi \approx 30\% \), while \( \pi \approx 40\% \) when \( 1 - \alpha = 10 \) (see again Appendix \ref{app:assetpricing}). In other words, an investor with such preferences would be willing to give up  either $30\%$ or $40\%$ of their wealth to remove persistence of consumption.   Since in the long-run risk model the persistent component is small, it seems unreasonable at the level of introspection to have such a high timing premium. To better understand whether this intuition is correct, I will now examine what reasonable levels of the persistence premium are supported by the experimental evidence on correlation aversion.

\vspace{0.15cm}
\noindent
\textbf{Measuring risk aversion via correlation aversion}. \cite{rohde2024intertemporal} propose  a model-free method to measure  correlation aversion.  Their approach quantifies the  degree of positive correlation aversion  \citep[p. 3496]{rohde2024intertemporal} by computing  the difference in present certainty
equivalents between an iid temporal lottery $d^{iid}$ and a perfectly positively correlated temporal lottery $d^{corr}$ relative to the present certainty
equivalent of the iid lottery, that is:
\begin{equation}\label{eq:measurerohde}
\Delta_{\text {POS }}^{\%}:=\frac{u^{-1}(V_0(d^{iid}))-u^{-1}(V_0(d^{corr}))}{u^{-1}(V_0(d^{iid}))}=1-\frac{u^{-1}(V_0(d^{corr}))}{u^{-1}(V_0(d^{iid}))}.
\end{equation}
Assuming that \(\beta = 0.998\) and \(\rho = 1/3\)---standard parameter specifications in the macro-finance literature---I find that one can match the observed level of \(\Delta_{\text{POS}}^{\%}\) with a level of risk aversion such that \(1 - \alpha \approx 1.2\). When $\rho = 0$, the observed level of $\Delta_{\text{POS}}^{\%}$ can instead be matched with a level of risk aversion satisfying $1 - \alpha \approx 1$ (see Section \ref{sec:measuring}). These results are consistent with existing estimates, which find \(1 - \alpha\) to be between \(1\) and \(2\) (see, for example, the discussion on p. 154 in \citealt{mehra1985equity}). 

 So why is the persistence premium $\pi$ under long-run risk so high? \cite{rohde2024intertemporal} find that correlation aversion does not depend on preference for information. However, as shown previously, Epstein–Zin preferences cannot distinguish between risk aversion and preference for information.  Indeed, when relative risk aversion $1-\alpha$ increases, the measure of preference for information increases as well since   
\begin{equation}ER_\phi(x,y)=\frac{y(1-\frac{\alpha}{\rho})}{ x(\beta x+y)}.\end{equation}
Therefore, based on the approximation in equation \eqref{eq:appropremium}, one can see that in the EZ case, a higher relative risk aversion parameter, \(1 - \alpha\), leads to greater values of \(ER_\phi(x, x)\) and \(ER_\phi(y, y)\). This effect is amplified given that \(\alpha < 0\) and that the EIS is large in the long-run risk model.  This, in turn, causes the persistence premium to grow at a slower rate as the level of persistence increases. This result explains why  one may need very high levels of risk aversion  and persistence to achieve the degree of correlation aversion necessary to match the equity premium.



\vspace{0.15cm}
\noindent
\textbf{An extension of Epstein–Zin preferences}.    I introduce a new model where the parameter of risk aversion does not necessarily increase preference for information. With this model, one can match the elicited level of correlation aversion in the experiment of \cite{rohde2024intertemporal}  with much lower levels of preference for information.      This point is connected to the one raised by \cite{meyer2019generalized} who showed  that a generalization of  EZ  preferences can produce Sharpe ratios comparable to the empirical values using more realistic parameters.

 I consider the following  HARA risk adjustment factor $\phi_{\gamma,b}$  \citep[p. 389]{merton1971optimum} given by
\begin{equation}\label{def:haraprefs}
\phi_{\gamma,b}(x)=\frac{1-\gamma}{\gamma}\left(\frac{x}{1-\gamma}+b\right)^\gamma\text{ for every }x\in C,
\end{equation}
where    $0\neq \gamma<1$ and $b\geq 0$. 
Preferences $\succeq$ admit a HARA recursive representation if  they admit a   KP recursive representation $(\phi_{\gamma,b}, u, \beta)$. 

The risk adjustment factor of EZ preferences corresponds to the case   $b=0$ and $\gamma={\alpha}/{\rho}$.  Notice that this risk adjustment factor permits a partial separation between risk aversion and preference for information, meaning that  high levels of risk aversion can coexist with   either a high or low degree of preferences for early resolution of uncertainty. See Appendix \ref{haracalculations} for a formal discussion of these facts.   This reasoning motivates the following recursive representation.

In Appendix \ref{haracalculations}, I show that \(\phi_{\gamma,b}\) satisfies UPI. Consequently, by Theorem \ref{theo1}, correlation aversion corresponds to IRRA, which is equivalent to \(b \geq 0\). Moreover, consistent with the intuition discussed earlier, I show that  HARA recursive preferences can replicate the experimentally observed correlation aversion reported by \cite{rohde2024intertemporal}, using \textit{significantly lower---and thus more realistic---levels of risk aversion} compared to standard EZ parametrizations commonly employed in asset pricing. In short, with a more reasonable utility specification, bonds  can be attractive without assuming implausibly high risk aversion.

\subsection{Application:  income taxation and social mobility}\label{app:progtax}
In a setting of intergenerational mobility, where multiple dynasties care about both today's consumption and future generations, the trade-off between hedging and information becomes a trade-off between  social mobility and the predictability of income status—that is, the extent to which future positions are dictated by one's current place in the distribution \citep{shorrocks1978measurement}.  

In this setting,  correlation aversion has a significant impact on redistribution policies. Indeed,   I show that  redistribution policies---which resemble an ``inheritance'' tax based on historical family income---weaken persistent links between parental and child outcomes, increasing social mobility compared to standard preferences. The same result can also be achieved through alternative policies that reduce long-term consumption inequality, such as the redistribution of education expenditures.

More specifically,  I consider a modified version of  \citeauthor{benabou2002tax}'s (\citeyear{benabou2002tax}) dynamic model of optimal income taxation with the additional assumption that the innate ability shock can be persistent. In this model, progressive income taxation can serve as a welfare-enhancing tool due to imperfections in credit and insurance markets. However, redistribution introduces distortions in agents' effort or savings decisions, reflecting the classic trade-off between equity and efficiency. One might expect that, as the persistence of innate ability increases, the optimal progressive tax rate (i.e., the one maximizing steady-state aggregate welfare) would become more progressive in order to mitigate the heightened risks arising from imperfections in insurance markets.

Contrary to this intuition, under standard preferences, the  tax rate that maximizes steady-state welfare  remains around 33–35\%, regardless of the level of persistence of innate ability. However, with recursive utility, correlation aversion amplifies the impact of greater persistence, leading to a significant increase in the optimal progressive tax rate, rising from 45\% to 51\%. As a consequence, correlation aversion leads to higher social mobility (see equation \ref{eq:mobilitymeasure} and the related discussion).

\vspace{0.15cm}
\noindent
\textbf{The model}. Consider  \citeauthor{park2009genes}'s  (\citeyear{park2009genes})  modified version of \citeauthor{benabou2002tax}'s model. There is a continuum of infinitely-lived agents or dynasties, indexed by $i \in [0,1]$. In each period $t = 0,1,2,\dots$, agent $i$ chooses consumption $c_t^i$ and labor supply $l_t^i$ to maximize intertemporal utility $U_t^i$, defined recursively by:
\[
U_t^i = \max_{c_t^i,\, l_t^i} \exp\left\{(1-\beta)\left(\ln c_t^i - (l_t^i)^\eta\right) + \beta \ln\left[\mathbb{E}_t\left((U_{t+1}^i)^\gamma\right)\right]^{1/\gamma}\right\},
\]
subject to the constraints:
\begin{align}
y_t^i &= (h_t^i)^\lambda (l_t^i)^\mu, \tag{i}\\[6pt]
\hat{y}_t^i &= c_t^i + e_t^i, \tag{ii}\\[6pt]
h_{t+1}^i &= k\, \xi_{t+1} (h_t^i)^\alpha (e_t^i)^\rho. \tag{iii}
\end{align}
Therefore, each dynasty has EZ  preferences that satisfy correlation aversion. In this model, \({1}/{(\eta-1)}\) is the elasticity of labor supply and \(1-\gamma\) is relative risk aversion. The parameter \(k\) scales human-capital formation; \(\alpha\) is the elasticity with respect to parental human capital; and \(\lambda\) and \(\mu\) are the output shares of human capital and labor. Income (\(y_t^i\)) and disposable income (\(\hat{y}_t^i\)) depend on labor supply (\(l_t^i\)) and human capital (\(h_t^i\)). Human capital (\(h_{t+1}^i\)) is determined by the innate ability shock (\(\xi_{t+1}\)), parental human capital (\(h_t^i\), where \(\ln h_0^i \sim N(m_0, \Delta_0)\)), and investment in education (\(e_t^i\)). The expectation operator $\mathbb{E}_t$ is conditional on  the realized human capital $h_t^i$.

 The constraints require that  (i) the  income of each generation is produced by combining inherited human capital with labor supply, (ii) that entire income must be allocated between consumption and educational investment in the next generation, and (iii) the child’s human capital next period arises from a technology that mixes parental human capital, the education investment just made, and an uninsurable innate ability shock.

 The innate ability shock $\xi_t$ can be interpreted as reflecting among other things cognitive ability, and evolves according to the relationship $$\log (\xi_t)=\phi \log (\xi_{t-1})+\varepsilon_t \quad \text { where } \varepsilon_t \sim N\left(\mu_{\varepsilon}, \sigma_{\varepsilon}^2\right),$$
where $\phi$ denotes the level of \textit{persistence} of innate ability.  \cite{durlauf1996theory} provides a theoretical foundation for persistence in innate ability.

The break-even level income  $\tilde{y}_t$ is defined implicitly by the balanced-budget constraint:
$$\int_0^1\left(y_t^i\right)^{1-\tau}\left(\tilde{y}_t\right)^\tau d i=\int_0^1 \left(h_t^i\right)^\lambda\left(l_t^i\right)^\mu d i,$$ and the disposable income  $\hat{y}_t^i$  is a loglinear function of market income,
$$
\hat{y}_t^i=\left(y_t^i\right)^{1-\tau}\left(\tilde{y}_t\right)^\tau,
$$ 
 where the elasticity $\tau$ measures the rate of progressivity of fiscal policy.  

The taxation mechanism operates as follows: after determining the market income for all agents, the government calculates the break-even income level, $\tilde{y}_t$. Agents then report their market income to the tax agency. If an agent's income, $y_t^i$, exceeds $\tilde{y}_t$, they pay a positive tax; otherwise, they receive a subsidy. The elasticity, $\tau$, reflects the progressivity of fiscal policy, with both average and marginal tax rates increasing when $\tau > 0$.

The planner's objective is to optimize the steady-state aggregate welfare  given by $\lim_{t\rightarrow\infty} W_t $, where $W_t= \int_0^1 \ln U_t^i  di$. \cite{park2009genes} shows that under standard discounted expected utility (i.e., $1-\gamma=1$, which corresponds to logarithmic utility) the optimal progressive tax rate is 33\% with no persistence of innate ability ($\phi = 0$) and equal to 35\% when persistence increases to $\phi = 0.6$.  This result suggests that greater persistence has minimal impact on redistribution through progressive taxation. However, I show that with  correlation-averse  preferences satisfying $1-\gamma=10$ (see Appendix \ref{calc:tax}) one has significantly higher variability: the optimal progressive tax rate is approximately $\tau^* = 45.25\%$ for $\phi = 0$ and $\tau^* = 51.72\%$ for $\phi = 0.6$. 

Notably, correlation aversion increases social mobility. To see this point formally, observe that $\log(h_t)^i$ follows an AR(2) process with autoregressive coefficients
$ (\alpha+\lambda \beta(1-\tau))+\phi$ and $-(\alpha+\lambda \beta(1-\tau)) \phi$. As an inverse measure of social mobility, we can therefore take the sum of these two coefficients:
\begin{equation}\label{eq:mobilitymeasure}
(\alpha + \lambda \beta (1 - \tau))\,(1 - \phi) + \phi.
\end{equation}
Because correlation-averse preferences lower this inverse measure relative to the standard case, they imply higher social mobility (see Appendix \ref{calc:tax} for details.)

This result relates to \citeauthor{gottschalk2002evaluation}'s (\citeyear{gottschalk2002evaluation}) argument that employing recursive utility  generates a preference for social mobility. Nevertheless, in this model society aggregates utility additively, so the effect depends not on the welfare criterion but rather on individual preferences, which exhibit correlation aversion and favor mobility over the predictability of income status.

\section{Concluding remarks and discussion} \label{implications}
This paper has explored the relationship between non-instrumental information and intertemporal hedging within  the framework of recursive preferences. I have shown  that under reasonable constraints on risk attitudes, a decision maker will value intertemporal hedging  more  than early resolution of uncertainty. I highlighted the importance of this trade-off in applications such as asset pricing and intergenerational mobility. 

A limitation of existing models is that  both correlation aversion and preference for information are entirely determined by risk aversion. To address this issue, the paper proposed a generalization of Epstein–Zin preferences to partially separate risk aversion   and  preference for non-instrumental information. 

Further research is needed to understand the implications of more general recursive utility models. Below, I connect my findings to related literature in economics as potential avenues for further research.

\vspace{0.15cm}
\noindent
\textbf{Stochastic impatience}. \cite{dejarnette2020time} and \cite{dillenberger2020stochastic} study stochastic impatience, an axiom that extends impatience to risky environments. Like correlation aversion, stochastic impatience is a normatively desirable behavioral postulate.   They find that  EZ and HS models exhibit stochastic impatience provided that the level of risk aversion is not excessively high relative to the inverse of the EIS parameter. The relationship between correlation aversion and stochastic impatience is represented in Figure \ref{stochasticimpatience}. In particular, correlation aversion can be compatible with stochastic impatience. Similar to my findings, their results also advocate for a more general specification of preferences in order to reduce the level of risk aversion used in applications.

\vspace{0.15cm}
\begin{figure} 
\begin{center}
\begin{tikzpicture}
\filldraw[gray,ultra thick,fill=white] (-3.2,-2.5) rectangle (6.8,4.5);
\scope 
\clip (0,0) circle (0);
\fill[white] (1,0) circle (1);
\endscope
\draw[brown,ultra thick] (0.5,1) circle (3);
\draw[color=blue!100,ultra thick]      (0.5,1) circle (2);
\draw[color=black!60!green,ultra thick]      (0.5,1.9) circle (0.75);
\draw[purple,ultra thick]      (0.5,0.1) circle (0.75);
\draw[color=black!60!yellow,ultra thick]      (3.2,1.05) ellipse (2.5cm and 2.3cm);
      \draw[color=blue!100,ultra thick] (-1,1.2) node[] { \text{CA}};
      \draw[brown,ultra thick] (-2,1.2) node[] { \text{IH}};
      \draw[color=black!60!yellow,ultra thick] (4.5,1.2) node[] { \text{SI}};
      \draw[purple,ultra thick] (0.5,0) node[] { \text{HS}};
      \draw[color=black!60!green,ultra thick] (0.5,2) node[] { \text{EZ}};
       \draw (5.5,4) node[color=gray] { \text{KP}};
\end{tikzpicture}
\end{center}
\caption{Relationship between correlation-averse (CA) preferences and  recursive preferences that satisfy intertemporal-hedging (IH), Epstein–Zin (EZ) preferences, and  stochastic impatience (SI)}
\label{stochasticimpatience}
\end{figure}

\noindent\textbf{Climate policy}.\label{climatepolicy}
\cite{cai2019social} develop a dynamic stochastic general equilibrium
 model to estimate the effect of economic and climate
risks on the social
cost of carbon (SCC). They consider productivity shocks that exhibit persistence, leading to consumption growth rates that display long-run risk as in \eqref{LRR}.  Combined with Epstein–Zin preferences, the inclusion of persistent productivity shocks  results in  substantially higher social cost of carbon compared to scenarios without productivity shocks (see pp. 2705-2706 in \citealt{cai2019social}). The persistence premium developed in this model can be used to quantify the cost of long-run climate risks.\vspace{0.15cm}

\noindent
\textbf{Utility smoothing and fiscal hedging}.\label{fiscalhedging}
\citeauthor{karantounias2018optimal} (\citeyear{karantounias2018optimal}, \citeyear{karantounias2022general}) shows that Epstein–Zin recursive preferences significantly alter standard Ramsey tax-smoothing policies. The planner adopts fiscal hedging, taxing less during downturns and more during upturns to mitigate income shocks, driven by aversion to volatility in future utilities (\cite{karantounias2018optimal}, p. 2284).


Such a feature of preferences emerges \textit{in spite} of the fact that recursive preferences value  early resolution of uncertainty. Instead,  this feature emerges from correlation aversion. As shown by Theorems \ref{theo1} and \ref{theo2}, aversion to volatility in future utilities—mathematically reflected by concavity of the certainty equivalent—is characterized by bounds on preferences for early resolution of uncertainty.  The findings of my paper demonstrate that the same implications for optimal fiscal policy may not hold when using recursive preferences that do not satisfy correlation aversion, as is the case with preferences that exhibit DRRA.

\appendix
\section{Additional Material and Proofs}
\subsection{Acronyms, Notation, and Technical Definitions}\label{sec:acronyms}
\begin{table}[h!]
\centering
\caption{List of Acronyms and Symbols}
\begin{tabular}{|l|l|}
\hline
\textbf{Acronym/Symbol} & \textbf{Description} \\
\hline
$V_t$                 & Recursive utility representation at time $t$ \\
$u$                & Utility function capturing intertemporal substitution \\
$\phi$                & Risk adjustment factor \\
$\beta$               & Discount factor \\
EIS                  & Elasticity of Intertemporal Substitution \\
IRRA                  & Increasing Relative Risk Aversion \\
DARA                  & Decreasing Absolute Risk Aversion \\
CRRA                  & Constant Relative Risk Aversion \\
HARA                  & Hyperbolic Absolute Risk Aversion \\
PERU                  & Preference for information/early resolution of uncertainty \\
UPI                  & Uniform preference for information \\
SCA                  & Strong correlation aversion \\
KP                    & Kreps-Porteus preferences \\
EZ                    & Epstein–Zin preferences \\
HS                    & Hansen and Sargent's multiplier preferences \\
$\pi(\varepsilon)$       & Persistence premium given a level of persistence $\varepsilon\in [0,1]$ \\
$\Delta_{\text {POS }}^{\%}$ & Measure of correlation aversion from \cite{rohde2024intertemporal} \\
${C}$         & Consumption set: $[0,\infty)$ or $(0,\infty)$ \\
$\Delta_b(X),\Delta_{s}(X)$         &  Borel probability measures with bounded support on $X$\\
${D}_{t,s}$       & Simple (finite support) temporal lotteries  at time $t$ \\
$D_{0,s}^*$     & Set of lotteries with defined correlation structures \\
$T$                   & Finite time horizon, with $T=2$  in Section \ref{attcorr} \\
\hline
\end{tabular}
\label{tab:acronyms-symbols}
\end{table}

\noindent
\textbf{Polish Spaces}. A \emph{Polish space} is a topological space that is:
\begin{enumerate}
    \item Completely metrizable (i.e., there exists a metric that induces the topology and makes the space complete), and
    \item Separably metrizable (i.e., there exists a countable, dense subset).
\end{enumerate}

\vspace{0.15cm}
\noindent
\textbf{Simple Probability Measures (\(\Delta_s(X)\)):}  
Let \(X\) be a Polish space. The space of \emph{simple probability measures} on \(X\), denoted \(\Delta_s(X)\), is the set of probability measures on \(X\) with \emph{finite support}. That is:
\begin{align*}
\Delta_s(X) = \big\{\mu \in \Delta(X) : \, &\text{there exists a finite subset } \{x_1, \dots, x_n\} \subseteq X \\
&\text{such that } \mu(\{x_1, \dots, x_n\}) = 1  \big\}.
\end{align*}

\vspace{0.15cm}
\noindent
\textbf{Borel Probability Measures with Bounded Support (\(\Delta_b(X)\)):}  
Let \(X\) be a Polish space. The space of \emph{Borel probability measures with bounded support} on \(X\), denoted \(\Delta_b(X)\), is the set of all Borel probability measures \(\mu\) on \(X\) such that the support of \(\mu\), \(\operatorname{supp}(\mu)\), is compact:
\[
\Delta_b(X) = \left\{\mu \in \Delta(X) : \operatorname{supp}(\mu) \text{ is compact} \right\}.
\]

\vspace{0.15cm}
\noindent
\textbf{Absolute Continuity (\(\ell \ll m\))}.  
Let \(\ell, m \in \Delta_b(X)\). The measure \(\ell\) is \emph{absolutely continuous} with respect to \(m\) (denoted \(\ell \ll m\)) if for every Borel set \(A \subseteq X\),
\[
m(A) = 0 \implies \ell(A) = 0.
\]

\vspace{0.15cm}
\noindent
\textbf{Radon-Nikodym Derivative (\(\frac{d\ell}{dm}\))}.  
If \(\ell \ll m\), the \emph{Radon-Nikodym derivative} \(\frac{d\ell}{dm}\) is a Borel-measurable function \(f : X \to \mathbb{R}_+\) such that:
\[
\ell(A) = \int_A f(x) \, dm(x), \quad \text{for all Borel sets } A \subseteq X.
\]

\vspace{0.15cm}
\noindent
\textbf{Weak* Topology}.  
The \emph{weak* topology} on $\Delta_b(X)$  is the coarsest topology such that for all continuous and bounded functions \(f: X \to \mathbb{R}\), the map:
\[
\mu \mapsto \int_X f(x) \, d\mu(x)
\]
is continuous. With this topology, $\mu_n$ converges to $\mu$ in the weak* topology if and only if:
\[
\int_X f(x) \, d\mu_n(x) \to \int_X f(x) \, d\mu(x) \quad \text{for all such } f.
\]
\subsection{Measuring preference for information}\label{subsec:measuringperu}
To measure attitudes toward early resolution, I introduce  the following notion of \textit{early resolution premium}. This notion quantifies how much a DM is willing to pay to have risk resolve at $t=1$ rather than gradually. Assume for simplicity that $T=2$,  and consider the temporal lotteries   given $c_0,k>0,x>y> 0$ 
  $$d^{early}(\pi)=\left(c_0,\frac{1}{2}\left(k(1-\pi),x(1-\pi)\right)\oplus \frac{1}{2}\left(k(1-\pi),y(1-\pi)\right)\right),$$
  and
 \[
d^{{gradual}}(\varepsilon) = 
\left( c_0, \frac{1}{2}\left(k,\left(\left(\frac{1}{2}+\frac{\varepsilon}{2}\right)x \oplus \left(\frac{1}{2}-\frac{\varepsilon}{2}\right)y\right)\right) \right.
\]
\[
\left. \oplus \frac{1}{2}\left(k,\left(\left(\frac{1}{2}-\frac{\varepsilon}{2}\right)x \oplus \left(\frac{1}{2}+\frac{\varepsilon}{2}\right)y\right)\right)\right).
\] 
 In words, the gradual lottery resolves late when $\varepsilon=0$ and early when $\varepsilon=1$. In contrast, the early lottery resolves always early but the payoffs are discounted by a factor of $(1-\pi)$. See Figure \ref{premiumlotteries} for a graphical representation of these lotteries.

\begin{figure}
\begin{center}
\begin{tikzpicture}[scale=0.8,font=\footnotesize]
\tikzstyle{solid node}=[circle,draw,inner sep=1.5,fill=black]
\tikzstyle{hollow node}=[circle,draw,inner sep=1.5]
\tikzstyle{level 1}=[level distance=15mm,sibling distance=4cm]
\tikzstyle{level 2}=[level distance=15mm,sibling distance=1.5cm]
\tikzstyle{level 3}=[level distance=15mm,sibling distance=1cm]
\node(0)[solid node,label=above:{$c_0$}]{}
child{node[solid node,label=above left:{$k$}]{}
child{node[hollow node,label=below:{$x$}]{} edge from parent node[left]{$	1$}}
child{node[hollow node,label=below:{$y$}]{} edge from parent node[right]{$0$}}
edge from parent node[left,xshift=-5]{$\frac{1}{2}$}
}
child{node[solid node,label=above right:{$k$}]{}
child{node[hollow node,label=below:{$x$}]{} edge from parent node[left]{$0$}}
child{node[hollow node,label=below:{$y$}]{} edge from parent node[right]{$1$}}
edge from parent node[right,xshift=5]{$\frac{1}{2}$}
};
\end{tikzpicture}
\begin{tikzpicture}[scale=0.8,font=\footnotesize]
\tikzstyle{solid node}=[circle,draw,inner sep=1.5,fill=black]
\tikzstyle{hollow node}=[circle,draw,inner sep=1.5]
\tikzstyle{level 1}=[level distance=15mm,sibling distance=3.5cm]
\tikzstyle{level 2}=[level distance=15mm,sibling distance=1.5cm]
\tikzstyle{level 3}=[level distance=15mm,sibling distance=1cm]
\node(0)[solid node,label=above:{$c_0$}]{} 
child{node[solid node,label=above left:{$k$}]{} 
child{node[hollow node,label=below:{$x$}]{} edge from parent node[left]{$\frac{1}{2}+\varepsilon$}}
child{node[hollow node,label=below:{$y$}]{} edge from parent node[right]{$\frac{1}{2}-\varepsilon$}}
edge from parent node[left,xshift=-5]{$\frac{1}{2}$}
}
child{node[solid node,label=above right:{$k$}]{}
child{node[hollow node,label=below:{$x$}]{} edge from parent node[left]{$\frac{1}{2}-\varepsilon$}}
child{node[hollow node,label=below:{$y$}]{} edge from parent node[right]{$\frac{1}{2}+\varepsilon$}}
edge from parent node[right,xshift=5]{$\frac{1}{2}$}
};
\end{tikzpicture}
\end{center}
\caption{Probability tree representation of two temporal lotteries with $T=2$}
\label{premiumlotteries}
\end{figure}

Consider preferences $\succeq$ with KP representation $(\phi,u,\beta)$, where $\phi\in \mathcal{C}^2$  is concave and  satisfies UPI.  Moreover, I assume $u(x)=x$  to simplify calculations.   Because $\phi$ satisfies UPI, by Proposition \ref{peruchar} we have that $V_0\left(d^{early}\left(0\right)\right)\geq V_0\left(d^{gradual}\left(\varepsilon\right)\right)$, and as we increase $\pi$, $V_0(d^{early}(\pi))$  decreases, so that we can denote by $\pi(\varepsilon)$  the  unique solution to the equation $$V_0(d^{gradual}(\varepsilon))=V_0(d^{early}(\pi)).$$
The timing premium $\pi(\varepsilon)$ therefore quantifies how much one is willing to pay to have risk resolve at $t=1$ rather than more gradually as measured by the parameter $\varepsilon$.
  The next corollary  of Proposition \ref{peruchar} provides an approximation of $\pi(\varepsilon)$ near $\varepsilon=1$. 
 \begin{corollary}\label{corollaryperu}There exists a  constant $k_1>0$ such that as $\varepsilon\rightarrow 1$ 
 $$\pi(\varepsilon)=k_1\int_{y}^{x} \frac{\phi'(k+\beta z)}{\phi'(z)}ER_{\phi}(z,k)dz\left(1-\varepsilon\right)+\mathrm{o}\left(\varepsilon-1\right).$$
 \end{corollary} 
 \begin{proof}
 See the Appendix.
 \end{proof}
 Therefore, the premium is approximated by an average of the values of  ${ER}_{\phi}(z,k)$ over the interval $[y,x]$. As the measure increases, the premium increases.
 \subsection{Intertemporal hedging}
\label{negativecorrelation}
Here I discuss  the difference between my notion of correlation aversion and  \citeauthor{kochov2015time}'s (\citeyear{kochov2015time}) notion of intertemporal hedging.  Consider the temporal lotteries $d=(c_0,m),d'=(c_0,m')\in D_{0,s}$ where for some $x,y\in C$ we have $m_1^{\prime}(x)=m_1(x)=\frac{1}{2}$, $m_2(x|x)=m_2(y|y)=1$, and $m_2^{\prime}(y|x)=m_2^{\prime}(x|y)=1$. Figure \ref{Intertemporal} provides a graphical representation of these two lotteries. The lottery $d$ is obtained by applying an IECIT with $\varepsilon=\frac{1}{2}$. The lotteries $d$ and $d'$ have perfect positive and negative correlation, respectively.

 We can immediately see that $d \geq_B d'$ and $d' \geq_B d$, meaning that $d$ and $d'$ are equally informative. The strict preference for $d'$ over $d$, is referred to as correlation aversion by  \cite{bommier2007risk} and  \textit{intertemporal hedging} by \cite{kochov2015time}.  I adopt the latter terminology as it reflects the fact that with their being equally informative only hedging considerations affect the evaluations of these two lotteries. The next result demonstrates that intertemporal hedging is equivalent to the concavity of $\phi$ (i.e., risk aversion).

\begin{proposition}\label{intertemporalhedgingproposition}
Preferences $\succeq$ with KP representation $(\phi,u,\beta)$ exhibit intertemporal hedging if and only if  $\phi$ is concave.
\end{proposition}
\begin{proof}
Observe that intertemporal hedging  is equivalent to 
$$\frac{1}{2}\phi(u+\beta u)+\frac{1}{2}\phi(v+\beta v) \leq \frac{1}{2}\phi(v+\beta u)+\frac{1}{2}\phi(u+\beta v),$$
for every $u,v\in u(C)$.
Therefore, the statement follows by a straightforward application of Theorem 4(a) in \cite{epstein1980increasing}.
\end{proof}
\begin{figure}
\begin{center}
\begin{tikzpicture}[scale=1,font=\footnotesize]
\tikzstyle{solid node}=[circle,draw,inner sep=1.5,fill=black]
\tikzstyle{hollow node}=[circle,draw,inner sep=1.5]
\tikzstyle{level 1}=[level distance=15mm,sibling distance=3.5cm]
\tikzstyle{level 2}=[level distance=15mm,sibling distance=1.5cm]
\tikzstyle{level 3}=[level distance=15mm,sibling distance=1cm]
\node(0)[solid node,label=above:{$0$}]{}
child{node[solid node,label=above left:{$x$}]{}
child{node[hollow node,label=below:{$x$}]{} edge from parent node[left]{$1$}}
edge from parent node[left,xshift=-3]{$\frac{1}{2}$}
}
child{node[solid node,label=above right:{$y$}]{}
child{node[hollow node,label=below:{$y$}]{} edge from parent node[right]{$1$}}
edge from parent node[right,xshift=3]{$\frac{1}{2}$}
};
\end{tikzpicture}
\begin{tikzpicture}[scale=1,font=\footnotesize]
\tikzstyle{solid node}=[circle,draw,inner sep=1.5,fill=black]
\tikzstyle{hollow node}=[circle,draw,inner sep=1.5]
\tikzstyle{level 1}=[level distance=15mm,sibling distance=3.5cm]
\tikzstyle{level 2}=[level distance=15mm,sibling distance=1.5cm]
\tikzstyle{level 3}=[level distance=15mm,sibling distance=1cm] 
\node(0)[solid node,label=above:{$0$}]{}
child{node[solid node,label=above left:{$x$}]{}
child{node[hollow node,label=below:{$y$}]{} edge from parent node[left]{$1$}}
edge from parent node[left,xshift=-3]{$\frac{1}{2}$}
}
child{node[solid node,label=above right:{$y$}]{}
child{node[hollow node,label=below:{$x$}]{} edge from parent node[right]{$1$}}
edge from parent node[right,xshift=3]{$\frac{1}{2}$}
};
\end{tikzpicture} 
\end{center}
\caption{Negative vs positive correlation}\label{Intertemporal}
\end{figure}
Observe that under the assumptions of Theorem \ref{theo1}, correlation aversion implies that $\phi$ is concave. Hence, from Proposition \ref{intertemporalhedgingproposition} we can infer that under these assumptions correlation aversion  implies  intertemporal hedging.
\subsection{Extension to  lotteries with infinite support}
 Note that here we operate under the convention that if $Y\subseteq X$, then $\Delta_b(Y)\subseteq \Delta_b(X)$ by identifying each probability measure in $\Delta_b(Y)$ with the equivalent probability measure in $\Delta_b(X)$ that assigns probability 1 to $Y$.  Endow  $\Delta_b(X)$ with the weak$^*$ topology.

 The extension uses  the following notion a uniformly bounded sequence of temporal lotteries.
\begin{definition}[Uniformly bounded temporal lotteries] Say that a sequence  $(c,m_n)_{n=0}^\infty$ in $D_0$ is uniformly bounded if there exists a compact set $K\subseteq \mathbb{R}$ such that for some $K$ $$m_n\in \Delta_b(K\times \Delta_b(K))\text{ eventually.}$$ 
\end{definition}

The idea is that \( d \in D_0 \) is considered more correlated than \( d' \in D_0 \) if both can be approximated  by sequences of uniformly bounded simple lotteries \( (c,m_n)_{n=0}^\infty \) and \( (c,m_n')_{n=0}^\infty \), respectively, where each \( (c,m_n) \) is more correlated than \( (c,m_n') \). Correlation aversion then implies that \( d' \) is preferred to \( d \).
\begin{proposition}\label{prop:corrextension}
Assume that the preferences \( \succeq \) exhibit correlation aversion. Consider \( d, d' \in D_0 \) such that there exist uniformly bounded sequences  \( (c,m_n)_{n=0}^\infty \) and \( (c,m_n')_{n=0}^\infty \) in \( D_{0,s}^* \), and a sequence \( (\ell_n)_{n=0}^\infty \) in \( \Delta_s(C) \), satisfying:
\[
\lim_{n \to \infty} (c,m_n) = d, \quad \lim_{n \to \infty} (c,m_n') = d',
\]
and for every \( n \geq 0 \),
\[
 (c,m_n) \geq_C  (c,m_n') \geq_C d^{iid}(\ell_n).
\]
Then \( d' \succeq_0 d \).
\end{proposition}

\subsection{The persistence premium and long-run risk}\label{app:assetpricing}
We have that (see  \cite{epstein2014much}, pp. 2684-2685)
$$
\log U_{0}(d^{corr})=\log c_{0}+\frac{\beta}{1-\beta a} x_{0}+\frac{\beta}{1-\beta} m+\frac{\alpha}{2} \frac{\beta \sigma^{2}}{1-\beta}\left(1+\frac{\varphi^{2} \beta^{2}}{(1-\beta a)^{2}}\right),
$$
and
$$
\log U_{0}(d^{iid})=\log c_{0}+\beta x_{0}+\frac{\beta}{1-\beta} m+\frac{\alpha}{2} \frac{\beta \sigma^{2}}{1-\beta}\left(1+\varphi^{2} \beta^{2}\right).
$$
Therefore, we obtain
$$\pi=1-\frac{U_0(d^{corr})}{U_0(d^{iid})}=1-e^{\frac{\beta}{1-\beta a} x_{0}-\beta x_0+\frac{\alpha}{2} \frac{\beta \sigma^{2}}{1-\beta}\left(\frac{\varphi^{2} \beta^{2}}{(1-\beta a)^{2}}-\varphi^{2} \beta^{2}\right)}.$$

\begin{align*}
\pi &= 1-\exp \left(
\scalebox{0.8}{$-6.5 \times 0.998 \times \frac{0.0078^{2}}{2(1-0.998)} 
\left(0.044^2 \times \frac{0.998^{2}}{(1-0.998 \times 0.979)^{2}} - 0.044^{2} \times 0.998^{2}\right)$}
\right) \\
&\approx 0.302, \\
\pi &= 1-\exp \left(
\scalebox{0.8}{$-9 \times 0.998 \times \frac{0.0078^{2}}{2(1-0.998)} 
\left(0.044^2 \times \frac{0.998^{2}}{(1-0.998 \times 0.979)^{2}} - 0.044^{2} \times 0.998^{2}\right)$}
\right) \\
&\approx 0.393.
\end{align*}

Therefore, we have that $\pi\approx 30\%$ with $1-\alpha=7.5$ and $\pi \approx 40\%$ with $1-\alpha=10$. 

 Note that these results do not change significantly  if  one increases  the long-run volatility of the iid process to match the volatility of the persistent process. Consider for example the case $1 - \alpha = 7.5$. Observe that
$$\lim_{t\rightarrow\infty}\operatorname{Var}\left(\log \frac{c_{t+1}}{c_t}\right)=\sigma^2+\frac{\varphi^2 \sigma^2}{1-a^2},$$
so that by setting  $\sigma \approx 0.0079719$ we obtain that the two processes have the same long-run volatility:
$$0.0078^2+\frac{0.044^2 \times 0.0078^2}{1-0.979^2}=\sigma^2+0.044^2 \sigma^2.$$
With this level of persistence in the i.i.d. process, when \( 1 - \alpha = 7.5 \), we obtain the persistence premium:

\[
\begin{aligned} 
1 - \exp\bigg( 
    -\frac{6.5}{2} \times \frac{0.998 \times 0.0078^2}{1 - 0.998} 
    & \left(1 + 0.044^2 \times \frac{0.998^2}{(1 - 0.979 \times 0.998)^2}\right) \\
    + \frac{6.5}{2} \times \frac{0.998 \times 0.0079719^2}{1 - 0.998} 
    & \left(1 + 0.044^2 \times 0.998^2\right)
\bigg) 
\approx 0.299790.
\end{aligned}
\]
\subsection{Measuring risk aversion}\label{sec:measuring}
Recall that from equation \eqref{eq:measurerohde} we have that the measure of correlation aversion is given by $$\Delta_{P O S}^{\%}=1-\frac{u^{-1}(V_0(d^{corr}))}{u^{-1}(V_0(d^{iid}))}.$$
The lotteries  considered in \cite{rohde2024intertemporal}  feature risk at $t=0$. However, since  EZ preferences are stationary,  one can equivalently consider the pair of temporal lotteries given by
\[
d^{corr}= \scalebox{0.9}{$
\left(0, \frac{1}{2}\left( 10,  10  \right) \oplus \frac{1}{2} \left( 5, 5 \right) \right)$}.
\]
  
  and    
\[
d^{iid} = \left(0, \scalebox{0.85}{$\frac{1}{2} \left( 10, \left( \frac{1}{2} 10 \oplus  
\frac{1}{2} 5  \right) \right) \oplus \frac{1}{2} \left( 5, 
\left( \frac{1}{2} 10  \oplus \frac{1}{2} 5 \right) \right)$} \right).
\]
Here, I assume that $t=1$ equals 4 weeks, as in the first time frame considered by \cite{rohde2024intertemporal}. Since the time unit is 4 weeks, we can apply the monthly discount factor $\beta=0.998$ used in \cite{bansal2004risks}. In this case, we have
$$u^{-1}\left(V_0\left(d^{corr}\right)\right)=u^{-1}\left\{ 0.998 \phi^{-1}\left( \phi\left(u(5) + 0.998 u(5)\right) + \phi\left(u(10) + 0.998 u(10)\right) \right) \right\},$$
and
\begin{align*}
u^{-1}\left(V_0\left(d^{iid}\right)\right)=u^{-1}\left\{ 0.998 \phi^{-1}\left( \phi\left(u(5) + 0.998 \phi^{-1}\left( \frac{1}{2} \phi(u(5)) 
+ \frac{1}{2} \phi(u(10)) \right)\right) \right.\right. \\
\left.\left. + \phi\left(u(10) + 0.998 \phi^{-1}\left( \frac{1}{2} \phi(u(5)) 
+ \frac{1}{2} \phi(u(10)) \right)\right)\right) \right\}
\end{align*}
I consider the mean value $$ \Delta_{P O S}^{\%} = 0.008 $$ found in their experiment \citep[p. 55]{rohde2022intertemporal}. This value corresponds to the time-risk framing of their experiment, which encouraged subjects to consider correlation over time.

 When ${1}/{(1-\rho)}=1.5$---a common specification in the long-run risk literature \citep{bansal2004risks} we can match this level of correlation aversion by setting $\alpha=-{0.61}/{3}\approx -0.2$ since in this case we obtain $$1-\frac{u^{-1}(V_0(d^{corr}))}{u^{-1}(V_0(d^{iid}))}\approx 0.008.$$ 

 
 When $\rho\approx 0$, by setting  $\alpha \approx  -0.0345$ we obtain  
$$1-\frac{u^{-1}(V_0(d^{corr}))}{u^{-1}(V_0(d^{iid}))}\approx 1-\frac{\left(\frac{5^{1.998 \alpha}+10^{1.998 \alpha}}{2}\right)^{\frac{0.998}{\alpha}}}{\left(\frac{1}{2}\left(5^\alpha+10^\alpha\right)\right)^{\frac{1.996 }{\alpha}}}\approx 0.008.$$
 Note that when $\alpha\approx -0.0345$, we have that the persistence premium $\pi$ is close to zero since
\begin{footnotesize}
\begin{align*}
\pi=1 - \exp\left( (-0.0345)\times\frac{0.998 \times 0.0078^2}{2 \times (1 - 0.998)} \left(\frac{0.044^2 \times 0.998^2}{\left(1 - 0.998 \times 0.979\right)^2} - 0.044^2 \times 0.998^2\right) \right)\\\approx 0.0019,
\end{align*}\end{footnotesize} 
a more reasonable value in terms of introspection and consistent with the evidence in \cite{meissner2022measuring}. Indeed,  they show that 40\% of subjects have a zero timing premium.


\subsection{The persistence premium and HARA recursive preferences}\label{haracalculations}
First observe that for every $\beta\in (0,1]$ and  $x,y\geq 0$
\begin{align*}
-\frac{\phi_{\gamma,b}^{\prime \prime}(x)}{\phi_{\gamma,b}^{\prime}(x)} + \beta \frac{\phi_{\gamma,b}^{\prime \prime}(\beta x+y)}{\phi_{\gamma,b}^{\prime}(\beta x+y)} &= \left(\frac{1}{\frac{x}{1-\gamma}+b} - \beta\frac{1}{\frac{\beta x+y}{1-\gamma}+b}\right) \\
&= \left(\frac{1}{\frac{x}{1-\gamma}+b} - \frac{1}{\frac{x+\frac{y}{\beta}}{1-\gamma}+\frac{b}{\beta}}\right) \geq 0,
\end{align*}
 which implies that UPI is satisfied. Further, we have that for $x>0$
$$R_{\phi_{\gamma,b}}(x)=\frac{x}{\frac{x}{1-\gamma}+b}=\frac{1}{\frac{1}{1-\gamma}+\frac{b}{x}}.$$
so that IRRA is satisfied whenever $b\geq 0$. 
Now observe that in the EZ case $b=0$ so that:
$$-\frac{\phi_{\gamma,0}^{\prime \prime}(x)}{\phi_{\gamma,0}^{\prime}(x)}+\beta \frac{\phi_{\gamma,0}^{\prime \prime}(\beta x+y)}{\phi_{\gamma,0}^{\prime}(\beta x+y)}=\frac{(1-\gamma) y}{x(\beta x+y)}\geq 0,$$
which implies that if risk aversion goes to infinity, i.e., if $\gamma\rightarrow-\infty$ then
\begin{equation}\label{hararelation}
-\frac{\phi_{\gamma,0}^{\prime \prime}(x)}{\phi_{\gamma,0}^{\prime}(x)}+\beta \frac{\phi_{\gamma,0}^{\prime \prime}(\beta x+y)}{\phi_{\gamma,0}^{\prime}(\beta x+y)}\rightarrow +\infty.
\end{equation}
Finally, we  have that when $\beta$ is close to unity 
$$\lim _{\gamma \rightarrow-\infty}-\frac{\phi_{\gamma,0}^{\prime \prime}(x)}{\phi_{\gamma,0}^{\prime}(x)}+\beta \frac{\phi_{\gamma,0}^{\prime \prime}(\beta x+y)}{\phi_{\gamma,0}^{\prime}(\beta x+y)}=\frac{1-\beta}{b}\approx 0.$$
Hence,  high levels of risk aversion are compatible with a small demand for non-instrumental information if we assume ``large'' values of $b$, while by \eqref{hararelation} for $b\approx 0$ one can have high levels of risk aversion compatible with high demand of non-instrumental information.

 To illustrate, assume that $u(x)=3{x^{1/3}}$ and $\phi_{\gamma,b}$ with  $\gamma= -2$,  and  $b=0.72$. Consider the temporal lotteries from \cite{rohde2024intertemporal}
\[
d^{corr}= \scalebox{0.9}{$
\left(0, \frac{1}{2}\left( 10,  10  \right) \oplus \frac{1}{2} \left( 5, 5 \right) \right)$}.
\]
  
  and    
\[
d^{iid} = \left(0, \scalebox{0.85}{$\frac{1}{2} \left( 10, \left( \frac{1}{2} 10 \oplus  
\frac{1}{2} 5  \right) \right) \oplus \frac{1}{2} \left( 5, 
\left( \frac{1}{2} 10  \oplus \frac{1}{2} 5 \right) \right)$} \right).
\]

We obtain that
$$\Delta_{P O S}^{\%}=1-\frac{u^{-1}(V_0(d^{corr}))}{u^{-1}(V_0(d^{iid}))}\approx 0.0341.$$

If we assume that  $\gamma= -27$ and $b=0$, that is a standard EZ parametrization   in which $\gamma={\alpha}/{\rho}$, $1-\alpha=10$, ${1}/{(1-\rho)}=1.5$. In this case I obtain
$$\Delta_{P O S}^{\%}=1-\frac{u^{-1}(V_0(d^{corr}))}{u^{-1}(V_0(d^{iid}))}\approx 0.0341.$$

Hence, these different formulation of recursive HARA preferences and EZ preferences attain the same level of correlation aversion. However, in the former case, we obtain an average value of the measure of preference for information $ER_{\phi}(x,x)$ equal to
$$\int_{u(5)}^{u(10)}ER_{\phi}(x,x)dx=\int_{u(5)}^{u(10)}\left(\frac{1}{\frac{x}{3}+0.722}-\frac{1}{\frac{1}{3}\left(x+\frac{x}{0.998}\right)+\frac{0.722}{0.998}}\right)  d x=0.212242,$$  
as opposed to the EZ preferences:
$$\int_{u(5)}^{u(10)}ER_{\phi}(x,x)dx=\int_{u(5)}^{u(10)} \frac{(1+27) x}{x(0.998 x+x)} d x=3.23792,$$
and a level of relative risk aversion
$$\int_{u(5)}^{u(10)}R_{\phi}(x)dx=\int_{u(5)}^{u(10)}  \frac{1}{\left(\frac{1}{3}+\frac{0.722}{x}\right) 5}  d x= 0.581891,$$
as opposed to relative risk aversion of $1-(-27)=28$ under EZ preferences. Hence, recursive HARA preferences  achieve a comparable level of correlation aversion as EZ preferences under the standard parametrization, but they exhibit a much more limited level of relative risk aversion that is consistent with empirical evidence and a small preference for information.

\subsection{Income taxation and social mobility}\label{calc:tax}
The steady level of aggregate welfare is derived in  Appendix B in \cite{park2009genes} under the difference that $\rho$ and $\beta$ are switched in the present notation. Here we assume that $\sigma_\varepsilon^2=\omega^2$  and $\mu_\varepsilon=-{\omega^2}/{2}$. The values of the parameters are those in Table \ref{table:paramtax}. 
\begin{figure}[h]
    \centering
    \begin{minipage}[b]{0.48\textwidth}
        \centering
        \includegraphics[width=\textwidth]{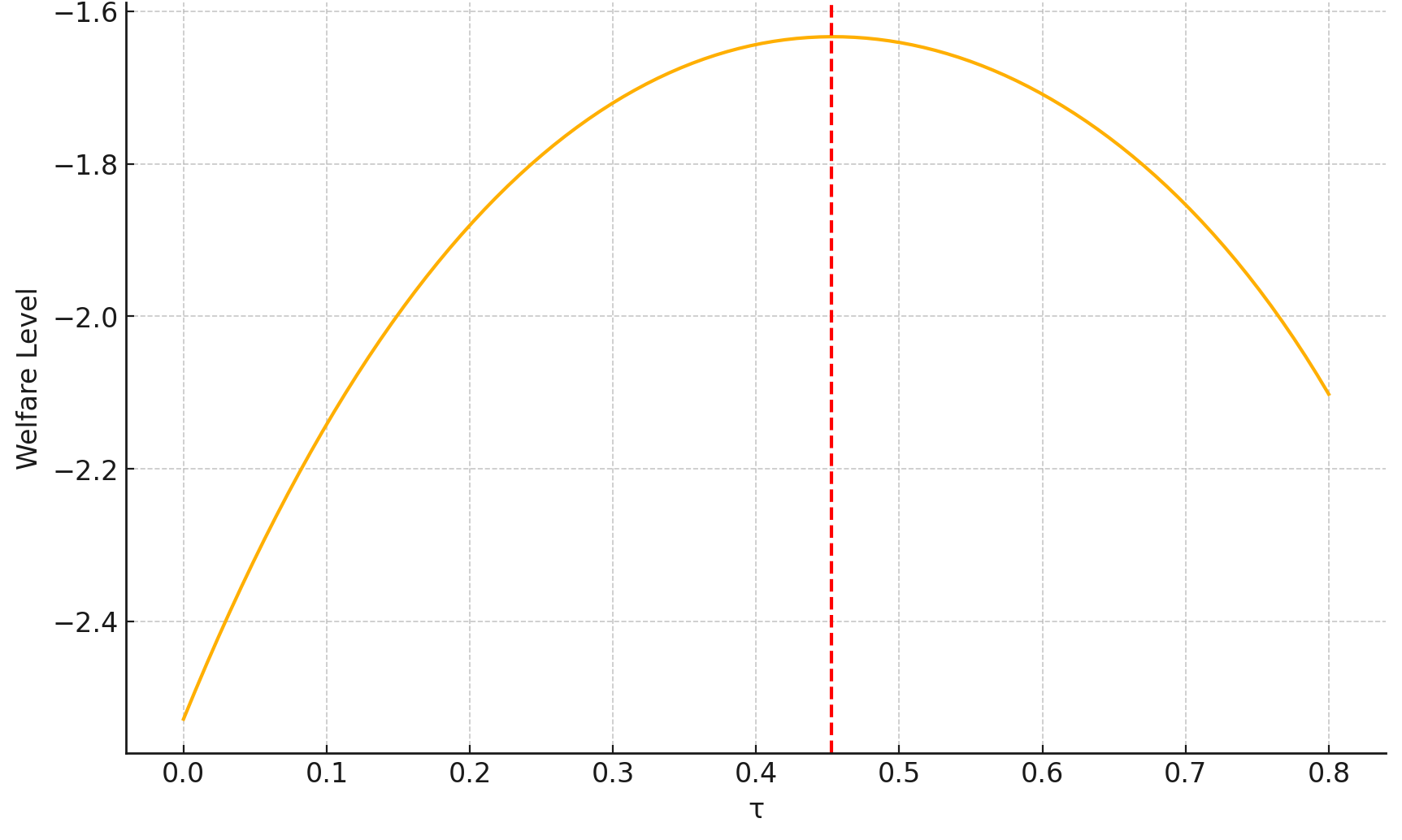}
        \caption{Optimal progressive tax  rate as a function of $\tau(\gamma=-9, \phi=0)$.}
        \label{fig:graphtax1}
    \end{minipage}
    \hfill
    \begin{minipage}[b]{0.48\textwidth}
        \centering
        \includegraphics[width=\textwidth]{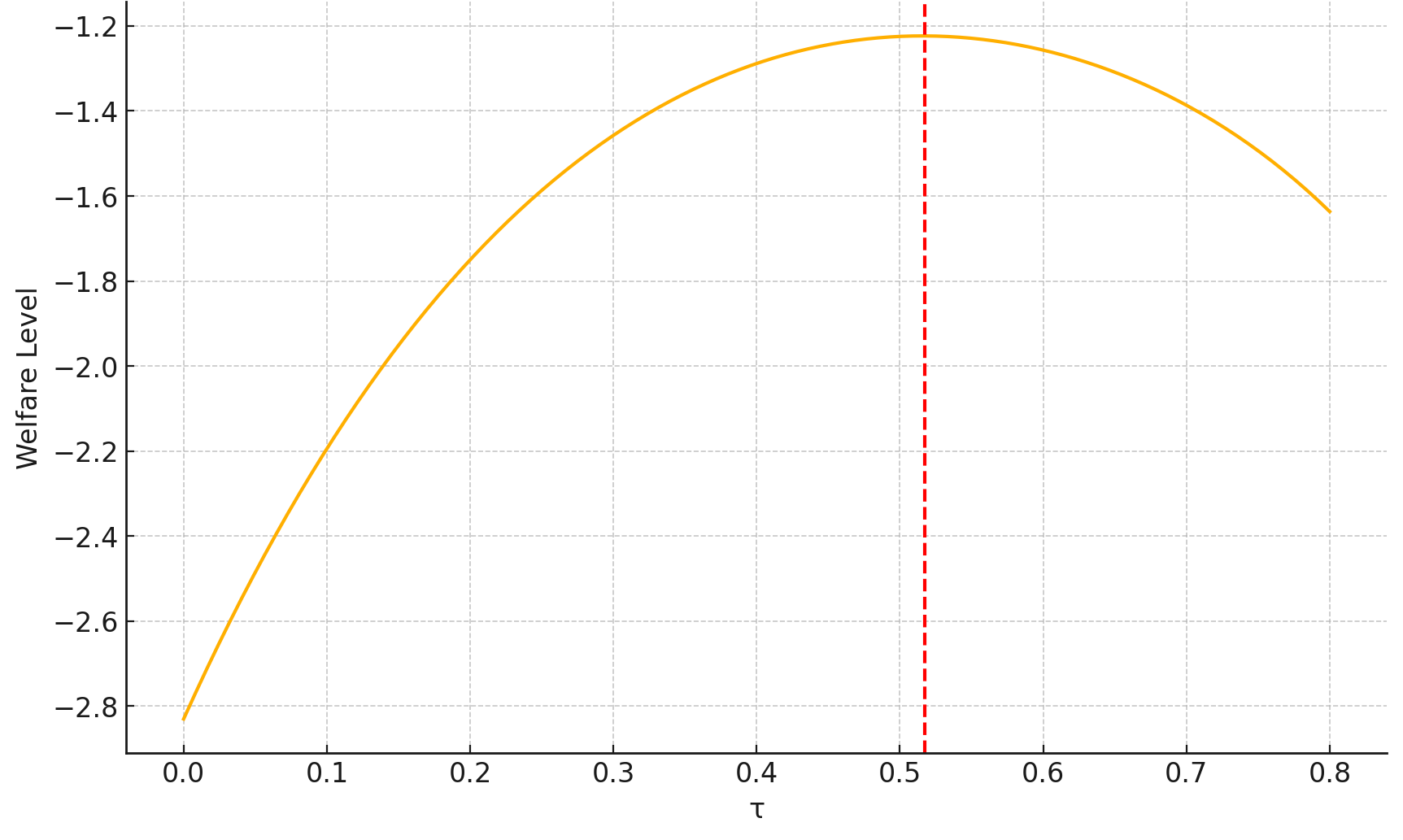}
        \caption{Optimal progressive tax rate as a function of $\tau(\gamma=-9, \phi=0.6)$.}
        \label{fig:graphtax2}
    \end{minipage}
\end{figure}

\begin{table}[h]
\centering
\caption{Parameters in the model}
\[
\begin{array}{|c|c|c|c|c|c|c|c|c|c|}
\hline
\beta & \gamma & \frac{1}{\eta - 1} & \phi & \lambda & \mu & \rho & \alpha & k & \omega \\
\hline
0.2939 & -9 & 0.2 & 0.6 & 0.625 & 0.375 & \frac{0.25}{0.625} & 0.35 & 1 & 1 \\
\hline
\end{array}
\]
\label{table:paramtax}
\end{table}


 The steady-state aggregate welfare as a function of $\tau\in [0,1]$  is given by 
\[
\begin{aligned}
& \frac{(1-\beta) \lambda \rho}{(1-\beta(\alpha+\rho \lambda))(1-\alpha-\rho \lambda)} \left( \ln (1-\tau) + \ln (\rho \beta \lambda) - \ln (1-\beta \alpha) + \frac{\mu}{\eta} \ln \left( \frac{\mu}{\eta} \right) + \right. \\
& \left. \frac{\mu}{\eta} \ln (1-\beta \alpha) - \frac{\mu}{\eta} \ln (1-\beta (\alpha + \rho \lambda (1-\tau))) + \frac{\tau(2-\tau) \lambda^2 \omega^2}{2 (1-\phi^2) (1 - \alpha - \rho \lambda + \rho \lambda \tau)^2} \right) \\
& + (1-\beta \alpha)(1-\beta(\alpha+\rho \lambda))^{-1} \left( \frac{\mu}{\eta} \ln \left( \frac{\mu}{\eta} \right) + \frac{\mu}{\eta} \ln (1-\beta \alpha) - \frac{\mu}{\eta} \ln (1-\beta (\alpha + \rho \lambda (1-\tau))) \right) \\
& - \frac{(\mu / \eta)(1-\beta \alpha)}{1-\beta(\alpha+\rho \lambda(1-\tau))} \\
& + \lambda \beta (1-\beta(\alpha+\rho \lambda))^{-1} \left( \ln k + \rho \ln (1-\tau) + \rho \ln (\rho \beta \lambda) - \rho \ln (1-\beta \alpha) \right) \\
& + \ln \left(1 - \frac{(1-\tau) \rho \beta \lambda}{1-\beta \alpha}\right) \\
& + \gamma \beta \frac{\lambda^2 (1-\beta) (1-\tau)^2 \omega^2}{2 (1-\phi^2) (1 - \beta \alpha - \beta \rho \lambda + \beta \rho \lambda \tau)^2} \\
& + (1-\beta \alpha)(1-\beta(\alpha+\rho \lambda))^{-1} \tau(2-\tau) \frac{\lambda^2 \omega^2}{2 (1-\phi^2) (1 - \alpha - \rho \lambda + \rho \lambda \tau)^2}
\end{aligned}
\]
The optimal $\tau^*\in [0,1]$  maximizes the previous expression. When $\phi=0$, I obtain that $\tau\approx  0.4525$.  When $\phi=0.6$, I obtain that $\tau\approx 0.5172$ (see Figures \ref{fig:graphtax1} and \ref{fig:graphtax2}).

Finally, observe that  social mobility is higher under correlation-averse preferences, with $\tau\approx 0.5172$, rather than under correlation neutrality which implies $\tau\approx 0.35$ (\cite{park2009genes}). Indeed,  when
$\alpha=0.35,  \lambda=0.625,  \beta=0.2939,  \phi=0.6$  we obtain
$$(\alpha+\lambda \beta(1-0.5172))(1-\phi)+\phi  =0.775<0.7878=(\alpha+\lambda \beta(1-0.35))(1-\phi)+\phi.$$
Hence, social mobility is higher under correlation-averse preferences.
\subsection{Proofs}
\noindent
\textbf{Proof of Proposition \ref{peruchar}.}  The proof uses the function $U_1\colon \Delta_{s}(D_{2,s}) \to \mathbb{R}$, defined for a fixed $\bar c \in C$ by
\begin{equation}\label{convexfunction}
U_1(m)=\phi\left(u(\bar{c})+\beta \phi^{-1}\left(\mathbb{E}_{m} \phi(V_{2})\right)\right)\quad \text{for every } m\in \Delta_s(D_{2,s}).
\end{equation}
%

\begin{lemma}\label{rital}
The function $U_1$ defined in \eqref{convexfunction} is convex if and only  if \eqref{convexity} holds.
\end{lemma}
\begin{proof}
First I claim that  $U_1$   is convex if and only  if the function $\Phi:\phi(u(C))\rightarrow\mathbb{R}$ defined by $x\mapsto \phi(\bar{c}+\beta\phi^{-1}\left(x\right))$ is convex. To see this point, observe that for every fixed $\bar{c}\in C$ we have that
\begin{align*}
U_1(\alpha m+(1-\alpha)m')\leq \alpha U_1(m)+(1-\alpha)U_1(m')\iff &   \\  \phi\left(\bar{c}+\beta\phi^{-1}\left(\alpha\mathbb{E}_m\phi\left(V_{2})+(1-\alpha)\mathbb{E}_{m^\prime}\phi(V_{2}\right)\right)\right)\leq \\ \alpha\phi\left(\bar{c}+\beta\phi^{-1}\left(\mathbb{E}_m\phi\left(V_{2}\right)\right)\right) +(1-\alpha)\phi\left(\bar{c}+\beta\phi^{-1}\left(\mathbb{E}_{m'}\phi\left(V_{2}\right)\right)\right)& .
\end{align*}
Since $u(C)$ is unbounded above and the statement above must hold  for every $m,m'\in\Delta_{s}(D_{2,s})$ it follows that convexity of $U_1$ is equivalent to $$\phi\left(\bar{c}+\beta\phi^{-1}\left(\alpha x+(1-\alpha)y\right)\right)\leq  \alpha\phi\left(\bar{c}+\beta\phi^{-1}\left(x \right)\right) +(1-\alpha)\phi\left(\bar{c}+\beta\phi^{-1}\left(y \right)\right),$$ for every $x,y\in \phi(u(C))$ which is equivalent to convexity of $\Phi$ for every $\bar{c}\in u(C)$.
Finally, the claim follows by using Lemma 3 in \cite{strzalecki2013temporal}.
\end{proof}

\begin{lemma}\label{lemma:tedeum} It holds that $$d_{KP}^{early} \geq_B d_{KP}^{late}.$$
\end{lemma}
\begin{proof}
First observe that since 
$$d_{KP}^{early}=\left(c_0, \bigoplus_{i=1}^n \pi_i\left(c_1, c_{2 i}, m_i\right)\right) \text{ and } \left(c_0, c_1, \bigoplus_{i=1}^n \pi_i\left(c_{2 i}, m_i\right)\right)= d_{KP}^{late},$$
we have that $\operatorname{marg}_{C} \bigoplus_{i=1}^n \pi_i\left(c_1, c_{2 i}, m_i\right)=\delta_{c_1}=\operatorname{marg}_{C} \left(c_1, \bigoplus_{i=1}^n \pi_i\left(c_{2 i}, m_i\right)\right)$.  Furthermore, the lotteries  $$\operatorname{marg}_{\Delta_s(D_{2,s})}\left(c_1, \bigoplus_{i=1}^n \pi_i\left(c_{2 i}, m_i\right)\right)=\delta_{\bigoplus_{i=1}^n \pi_i\left(c_{2 i}, m_i\right)}\in \Delta_s\left(\Delta_s\left(D_{2,s}\right)\right),$$ and $$\operatorname{marg}_{\Delta_s(D_{2,s})}\bigoplus_{i=1}^n \pi_i\left(c_1, c_{2 i}, m_i\right)=\bigoplus_{i=1}^n \pi_i\delta_{\left(c_{2 i}, m_i\right)}\in \Delta_s\left(\Delta_s\left(D_{2,s}\right)\right),$$  can be associated with  the matrix-vector pairs
\[
M\left[\operatorname{marg}_{\Delta_s(D_{2,s})}  \delta_{\bigoplus_{i=1}^n \pi_i\left(c_{2 i}, m_i\right)}\right]=\begin{bmatrix}
    \pi_1 & \ldots  & \pi_n  \\
\end{bmatrix},
\]    $M\left[\operatorname{marg}_{\Delta_s(D_{2,s})}\bigoplus_{i=1}^n \pi_i\delta_{\left(c_{2 i}, m_i\right)}\right]=I$, where $I$ denotes the identity matrix, $$\mu\left[\operatorname{marg}_{\Delta_s(D_{2,s})}  \delta_{\bigoplus_{i=1}^n \pi_i\left(c_{2 i}, m_i\right)}\right]=\left[1\right]\text{ and }\mu\left[\bigoplus_{i=1}^n \pi_i\delta_{\left(c_{2 i}, m_i\right)}\right]=\left[ \pi_1 \ldots \pi_n \right].$$
Now by setting $G:=\begin{bmatrix}
    \pi_1  & \ldots & \pi_n 
\end{bmatrix}$
we obtain that $$M\left[\operatorname{marg}_{\Delta_s(D_{2,s})}  \delta_{\bigoplus_{i=1}^n \pi_i\left(c_{2 i}, m_i\right)}\right]=\begin{bmatrix}
    \pi_1 & \ldots  & \pi_n 
\end{bmatrix}I=GM\left[\operatorname{marg}_{\Delta_s(D_{2,s})}\bigoplus_{i=1}^n \pi_i\delta_{\left(c_{2 i}, m_i\right)}\right],$$
and
 $$\mu\left[\operatorname{marg}_{\Delta_s(D_{2,s})}  \delta_{\bigoplus_{i=1}^n \pi_i\left(c_{2 i}, m_i\right)}\right]G=\left[1\right]\begin{bmatrix}
    \pi_1 & \ldots  & \pi_n 
\end{bmatrix}=\left[ \pi_1 \ldots \pi_n \right]=\mu\left[\bigoplus_{i=1}^n \pi_i\delta_{\left(c_{2 i}, m_i\right)}\right].$$ 
Hence, we can conclude that $d_{KP}^{early} \geq_B d_{KP}^{late}$ as desired.
\end{proof}
\begin{proof}[Proof of Proposition \ref{peruchar}] 
Note that if $\phi$ satisfies \eqref{convexity}, then each $U_t$ is convex by Lemma \ref{rital}. By Lemma \ref{lemma:tedeum},  $\operatorname{marg}_{\Delta_s(D_{2,s})}\left(c_1, \bigoplus_{i=1}^n \pi_i\left(c_{2 i}, m_i\right)\right)=\delta_{\bigoplus_{i=1}^n \pi_i\left(c_{2 i}, m_i\right)}$  is a garbling of $\operatorname{marg}_{\Delta_s(D_{2,s})}\bigoplus_{i=1}^n \pi_i\left(c_1, c_{2 i}, m_i\right)=\bigoplus_{i=1}^n \pi_i\delta_{\left(c_{2 i}, m_i\right)}$. By Theorem 4 in \cite{kihlstrom1984bayesian}, it follows that $$\sum_{i=1}^n \pi_i W\left({\left(c_{2 i}, m_i\right)}\right)\geq    W\left({\bigoplus_{i=1}^n \pi_i\left(c_{2 i}, m_i\right)}\right),$$ for every  convex function $W:\Delta_s(D_{2,s})\rightarrow\mathbb{R}$. 
 
Observe that \begin{align*}
V_0\left(d_{KP}^{early}\right) &\geq V_0\left(d_{KP}^{late}\right) \iff \\
\sum_{i=1}^n \pi_i \phi\left(u(c_1) + \beta V_{2}\left(c_{2i}, m_i\right)\right) &\geq \phi\left(u(c_1) + \beta \phi^{-1}\left(\sum_{i=1}^n \pi_i \phi \left(V_{2}\left(c_{2i}, m_i\right)\right)\right)\right).
\end{align*} 
 Since 
$$
\sum_{i=1}^n \pi_i U_1\left( {(c_{2 i}, m_i)} \right)= \sum_{i=1}^n \pi_i \phi\left( u(c_1) + \beta V_{2}( (c_{2 i}, m_i) ) \right)  
$$
and
$$
 U_1\left( {\bigoplus_{i=1}^n \pi_i (c_{2 i}, m_i)} \right)=\phi\left( u(c_1) + \beta \phi^{-1}\left( \sum_{i=1}^n \pi_i \phi \left( V_{2}( (c_{2 i}, m_i) ) \right) \right) \right),
$$
which by convexity of $U_1$  we obtain $\sum_{i=1}^n \pi_i U_1\left( (c_{2 i}, m_i) \right)\geq U_1\left(\bigoplus_{i=1}^n \pi_i (c_{2 i}, m_i) \right)$, and therefore that $d_{KP}^{early} \succeq_0 d_{KP}^{late}$.

 Conversely, consider $d,d'\in D_{0,s}$ given by
$$
d=\left(c_0,  \alpha\left(\bar{c}, m_1\right)\oplus (1-\alpha)\left(\bar{c}, m_2\right)\right),
$$ 
and
$$d'=\left(c_0, \bar{c}, \alpha m_1\oplus (1-\alpha)m_2\right),$$
 where $\alpha\in [0,1]$ and $V_2(m_1)=x$, $V_2(m_2)=y$. We have that $d\succeq_0 d'$ if and only if
 $$ \alpha\phi(\bar{c}+\beta\phi^{-1}(x))+(1-\alpha)\phi(\bar{c}+\beta\phi^{-1}(y))\geq \phi(\bar{c}+\beta\phi^{-1}(\alpha x+(1-\alpha y))).$$
 Since the statement has to hold for arbitrary $x,y\in u(C)$ (recall that $u$ is unbounded above) and $\alpha\in [0,1]$, it follows that the mapping $x\mapsto \phi(\bar{c}+\beta\phi^{-1}\left(x\right))$ must be convex. Hence an immediate application of Lemma \ref{rital} concludes the proof.
 \end{proof}

\noindent
\textbf{Proof of Proposition \ref{iidinfo}.} 
\begin{proof}
Denote by $\{c,c',\ldots,c_N\}$ the support of $\ell\in\Delta_s(C)$. It suffices to show that if $d'\in D_{0,s}^*$ differs from some $d^{iid}(\ell)\in D_{0,s}^*$ by an IECIT and $d\in D_{0,s}^*$ differs from $d'$ by an IECIT then  $d\geq_B d'\geq_B d^{iid}(\ell)$. Suppose that $d'$  differs from  $d^{iid}(\ell)$ by an IECIT and  that $d$ differs from $d'$ by an IECIT. First observe that $d=(c_0,m),d^{\prime}=(c_0,m')$ where $\operatorname{marg}_{C} m'=\operatorname{marg}_{C} m$ since by definition of an IECIT $m_1=m_1'$.  Now consider the stochastic matrices of  conditional distributions for $d$ and $d'$:
 \[
A:=M\left[\operatorname{marg}_{\Delta_s(C)} m\right]=\begin{bmatrix}
    m_2(c|c)       & m_2(c'|c) & \ldots  & m_2(c_N|c) \\
  m_2(c|c')       & m_2(c'|c') & \ldots & \vdots\\
     \vdots           &     \vdots    & \ddots &  \vdots \\ 
      m_2(c|c_N)          &     m_2(c'|c_N)    & \ldots &  m_2(c_N|c_N) \\ 
\end{bmatrix},
\]
and
 \[B:=M\left[\operatorname{marg}_{\Delta_s(C)} m'\right]=\begin{bmatrix}
    m_2^\prime(c|c)       & m_2^\prime(c'|c) & \ldots  & m_2'(c_N|c) \\
  m_2^\prime(c|c')       & m_2^\prime(c'|c') & \ldots & \vdots\\
     \vdots           &     \vdots    & \ddots &  \vdots \\ 
      m_2^\prime(c|c_N)          &     m_2^\prime(c'|c_N)    & \ldots &  m_2^\prime(c_N|c_N) \\ 
\end{bmatrix}.
\]
  Then for some $\varepsilon,\varepsilon'\geq 0$ 
  \[
\begin{bmatrix}
    m_2(c|c)       & m_2(c'|c)  \\
  m_2(c|c')       & m_2(c'|c')  
\end{bmatrix}=\begin{bmatrix}
    m_2'(c|c)+\frac{\varepsilon}{\ell(c)}       & m_2'(c'|c)-\frac{\varepsilon}{\ell(c)}  \\
  m_2'(c|c')-\frac{\varepsilon}{\ell(c')}     & m_2'(c'|c') +\frac{\varepsilon}{\ell(c')}
\end{bmatrix},
\]
and 
\[
 \begin{bmatrix}
    m_2'(c|c)      & m_2'(c'|c) \\
  m_2'(c|c')     & m_2'(c'|c')
\end{bmatrix}=\begin{bmatrix}
    \ell(c)+\frac{\varepsilon'}{\ell(c)}       & \ell(c')-\frac{\varepsilon'}{\ell(c)}  \\
  \ell(c)-\frac{\varepsilon'}{\ell(c')}     & \ell(c') +\frac{\varepsilon'}{\ell(c')}
\end{bmatrix}.
\]
  Hence, if we  choose $x_1,x_2\in[0,1]$ satisfying  $$x_1(\frac{\varepsilon+\varepsilon'}{\ell(c)})-(1-x_1)(\frac{\varepsilon+\varepsilon'}{\ell(c')})=\frac{\varepsilon'}{\ell(c)} \text{ and }x_2(\frac{\varepsilon+\varepsilon'}{\ell(c)})-(1-x_2)(\frac{\varepsilon+\varepsilon'}{\ell(c')})=\frac{\varepsilon'}{\ell(c')},$$ then letting
\[
G:=\begin{bmatrix}
    x_1 & 1-x_1 & 0  & \ldots & 0\\
    x_2 & 1-x_2 &  0 &  \ldots & 0\\
0       &  0        & 1 &   \ldots  & 0 \\
 \vdots   &  \vdots    & \vdots   & \ddots      \\
  0 & \vdots  & \vdots    & 0 &  1 \\
\end{bmatrix}
\]  
    it holds that
\[
B
=G A,
\] which implies that   $d\geq_B d'$. If instead we choose $$x_1=x_2=\frac{\frac{\ell(c)}{\ell(c')}}{1+\frac{\ell(c)}{\ell(c')}},$$ then  we obtain that
 \[\begin{bmatrix}
    \ell(c)      & \ell(c') & \ldots  & \ell(c_N) \\
  \ell(c)      & \ell(c') & \ldots & \vdots\\
     \vdots           &     \vdots    & \ddots &  \vdots \\ 
       \ell(c)         &      \ell(c') & \ldots &   \ell(c_N) \\ 
\end{bmatrix}=G B,
\]
which implies $d'\geq_B d^{iid}(\ell)$ as desired.
\end{proof}

\noindent
\textbf{Proof of Theorem \ref{theo1}.} Let  $d,d'\in D_{0,s}^*$ and $\ell\in \Delta_s(C)$.  I provide first the following preliminary result.
\begin{lemma}\label{brezislemma}Consider $d,d'$ such that $d'$ differs from some $d^{iid}(\ell)$ by an IECIT and $d$ differs from $d'$ by an IECIT.  Then there exists a twice continuously  differentiable function $U:[0,1]\rightarrow\mathbb{R}$ such that
\begin{enumerate}
\item $U(0)=\tilde{V}_0(d)$ and $U(1)=\tilde{V}_0(d')$;
\item $\lim_{\varepsilon\rightarrow 0}U'(\varepsilon)\leq 0$ whenever $d=d^{iid}(\ell)$;
\item  $U''(\varepsilon)\geq 0$ for every $\varepsilon\in (0,1)$.
\end{enumerate}
\end{lemma}
\begin{proof}
See Section \ref{proofs} in the Supplemental Appendix.
\end{proof}
It is now possible to prove Theorem \ref{theo1}. To this end, given $\ell\in\Delta_s(C)$, denote by $d^{corr}(\ell)=(c,m)\in D_{0,s}^*$ defined by $m_1=\ell$ and $m_2(c|c)=1$ for every $c\in \operatorname{supp} \ell$.
\begin{proof}[Proof of Theorem \ref{theo1}]
First, notice that for every $(c,m),(c,m')\in D_0$, it holds that $$V_0(c,m)\geq V_0(c,m')\iff \tilde{V}_0(c,m)\geq \tilde{V}_0(c,m').$$

 Assume now that $\phi$ satisfies IRRA and that $d\geq_C d'\geq_C d^{iid}(\ell)$. I claim that 
\[
d^{iid}(\ell) \succeq_0 d' \succeq_0 d,
\]
for every preference relation $\succeq$ with KP representation $(\phi,u,\beta)$. 

By Lemma \ref{brezislemma}, there exists a function ${U}:[0,1]\rightarrow\mathbb{R}$ such that, for some $q_1,q_2\in [0,1]$ with $q_1<q_2$, 
\[
U(0)=\tilde{V}_0(d^{iid}(\ell)),\quad U(q_1)=\tilde{V}_0(d'),\quad U(q_2)=\tilde{V}_0(d),\quad U(1)=\tilde{V}_0(d^{corr}(\ell)),
\]
$\lim_{\varepsilon\rightarrow 0} U'(\varepsilon)\leq 0$, and $U''(\varepsilon)\geq 0$ for almost every $\varepsilon\in (0,1)$. 

To see this, observe that given any sequence $(d_i)_{i=0}^N$ such that each $d_i$ differs from $d_{i-1}$ by an IECIT, one can repeatedly apply Lemma \ref{brezislemma} to construct a continuous function $U:[0,1]\rightarrow\mathbb{R}$ such that $U'(\varepsilon)$ and $U''(\varepsilon)$ exist for almost every $\varepsilon\in (0,1)$ and, moreover, the  limit $\lim_{\varepsilon\rightarrow 0}U'(\varepsilon)$ exists. Indeed, one first applies Lemma~\ref{brezislemma} to obtain twice continuously differentiable functions $(U_i)_{i=1}^N$ such that
\[
U_i(0)=\tilde{V}_0(d_{i-1})\quad\text{and}\quad U_i(1)=\tilde{V}_0(d_i)\quad\text{for every }i=1,\ldots,N,
\]
along with the other smoothness properties given in Lemma~\ref{brezislemma}. Then define $U$ by concatenating these functions on the subintervals of $[0,1]$: for $i=1,\ldots,N$ and $x\in\big[\frac{i-1}{N},\frac{i}{N}\big]$, set
\[
U(x):=U_i\left({Nx}-{(i-1)}\right).
\]
Each $U_i$ is twice continuously differentiable on $[0,1]$, hence Lipschitz on that interval. It follows that $U$ is piecewise Lipschitz continuous on $[0,1]$. By Lemma~2 in~\cite{Leobacher2022}, $U$ is therefore Lipschitz continuous on $[0,1]$ and therefore $U'$ exists for almost every $\varepsilon\in (0,1)$. Applying the same reasoning to $U'$ shows that $U''$ also exists for almost every $\varepsilon\in (0,1)$.

I now claim that $U$ satisfies
\[
\lim_{\varepsilon\rightarrow 1} U'(\varepsilon)\leq 0.
\]
Indeed, for some $p,q\in (0,1)$ and $x,y\in u(C)$ with $x>y$, we have
\[
\begin{aligned}
\lim_{\varepsilon\rightarrow 1} U'(\varepsilon)
= &\lim_{\varepsilon\rightarrow 1} \frac{\partial}{\partial \varepsilon} \Bigg[
p \,\phi\Big(x+\beta \phi^{-1}\big(\phi(x)(p+q\varepsilon)+\phi(y)(q-q\varepsilon)\big)\Big) \\
&\qquad\qquad\qquad\qquad +\;
q\,\phi\Big(y+\beta \phi^{-1}\big(\phi(x)(p-p\varepsilon)+\phi(y)(q+p\varepsilon )\big)\Big)
\Bigg]\\ 
\leq &\; \big(\phi(x)-\phi(y)\big)\left(\frac{\phi'((1+\beta)x)}{\phi'(x)}-\frac{\phi'((1+\beta)y)}{\phi'(y)}\right) \\
=&\; \big(\phi(x)-\phi(y)\big)\int_y^{x}\frac{(1+\beta)\,\frac{\phi^{\prime \prime}(z(1+\beta))}{\phi^{\prime}(z(1+\beta))}- \frac{\phi^{\prime \prime}(z)}{\phi^{\prime}(z)}}{\big(\phi'(z)\big)^2}\,dz \;\leq\; 0,
\end{aligned}
\]
where the last inequality follows from IRRA of $\phi$, upon observing that
\[
(1+\beta) \frac{\phi^{\prime \prime}(z(1+\beta))}{\phi^{\prime}(z(1+\beta))}- \frac{\phi^{\prime \prime}(z)}{\phi^{\prime}(z)}\leq 0 
\iff 
-z(1+\beta) \frac{\phi^{\prime \prime}(z(1+\beta))}{\phi^{\prime}(z(1+\beta))}\geq -z \frac{\phi^{\prime \prime}(z)}{\phi^{\prime}(z)}.
\]

Since $U$ and $U'$ are Lipschitz continuous, they are also absolutely continuous. Hence, for every $\tilde{\varepsilon}\in (0,1)$, since $U''\geq 0$ almost everywhere
\[
\lim_{\varepsilon\rightarrow 1}U'(\varepsilon)- U'(\tilde{\varepsilon})=\int_{\tilde{\varepsilon}}^1 U''(t)\,dt\geq 0.
\]
Combining this with $\lim_{\varepsilon\rightarrow 1} U'(\varepsilon)\leq 0$ implies that $U'(\tilde{\varepsilon})\leq 0$ for every $\tilde{\varepsilon}\in(0,1)$. Therefore
\[
\tilde{V}_0(d')-\tilde{V}_0(d^{iid}(\ell))
=\int_0^{q_1}U'(\tilde{\varepsilon})\,d\tilde{\varepsilon}\leq 0,
\]
and
\[
\tilde{V}_0(d')-\tilde{V}_0(d)
=\int_{q_1}^{q_2}U'(\tilde{\varepsilon})\,d\tilde{\varepsilon}\leq 0.
\]
Hence we obtain
\[
d^{iid}(\ell) \succeq_0 d' \succeq_0 d
\]
for every $\succeq$ with KP representation $(\phi,u,\beta)$, as desired.

As for the converse, assume that $\phi$ is not concave. Then there exist $x>y>0$ such that for every $z\in (y,x)$,
\begin{equation}\label{eq:notconcave}
\phi''\left(z(1+\beta)\right)>0.
\end{equation}
Consider the lottery $d^{iid}(\ell)$ with $\ell(x)=\ell(y)=\frac{1}{2}$. For each $\varepsilon\in [0,1]$, define $d^{\varepsilon}(\ell)=(c_0,m)$ by setting $m_2(x|x)=\ell(x)+\frac{1}{2}\varepsilon$ and $m_2(y|y)=\ell(y)+\frac{1}{2}\varepsilon$. Then, $d^{\varepsilon}(\ell)\geq_{C} d^{iid}(\ell)$. Define the function $U:[0,1]\rightarrow\mathbb{R}$ by $U(\varepsilon)=\tilde{V}_0(d^{\varepsilon}(\ell))$. Using equation \eqref{eq:notconcave} and the reasoning analogous to Lemma~\ref{brezislemma}, we obtain
\[
\lim_{\varepsilon\rightarrow 0} U'(\varepsilon)= \frac{(\phi(x)-\phi(y))}{\phi'\left(\phi^{-1}\left(\frac{\phi(x)+\phi(y)}{2}\right)\right)}\int_{y}^{x}\phi'(z)\phi''\left(z(1+\beta)\right) dz>0,
\]
which implies there exists $\bar{\varepsilon}\in(0,1)$ such that $U'(\tilde\varepsilon)>0$ for all $\tilde\varepsilon\in(0,\bar{\varepsilon})$. Consequently,
\[
\tilde{V}_0(d^{\bar{\varepsilon}}(\ell))-\tilde{V}_0(d^{iid}(\ell))=\int_{0}^{\bar{\varepsilon}}U'(\varepsilon)\,d\varepsilon>0,
\]
and thus $d^{\bar{\varepsilon}}(\ell)\geq_{C}d^{iid}(\ell)$ but $d^{\bar{\varepsilon}}(\ell)\succ_0 d^{iid}(\ell)$. Hence, $\phi$ must be concave.

Finally, suppose that $\phi$ does not satisfy IRRA. Since $\phi\in \mathcal{C}^3$, the function $R_{\phi}$ is continuously differentiable. Thus, there exist $\ubar{z}<\bar{z}$ in $\interior u(C)$ such that $R_{\phi}$ is non-increasing on the interval $[\ubar{z},\bar{z}]$ and $R_{\phi}(\bar{z})<R_{\phi}(\ubar{z})$. Choose $\beta\in(0,1]$ so that $\frac{\bar{z}}{1+\beta}>\ubar{z}$, and set $x={\bar{z}}/{(1+\beta)}$ and $y=\ubar{z}$. Consider again the lottery $d^{iid}(\ell)$ with $\ell(x)=\ell(y)={1}/{2}$, and define the lottery $d^{\varepsilon}(\ell)=(c_0,m)$ with $$m_2(x|x)=\ell(x)+\frac{1}{2}\varepsilon \text{ and }  m_2(y|y)=\ell(y)+\frac{1}{2}\varepsilon.$$ For $\varepsilon\geq \varepsilon'$, it holds that $d^{\varepsilon}(\ell)\geq_{C}d^{\varepsilon'}(\ell)\geq_C d^{iid}(\ell)$.

As before, define $U:[0,1]\rightarrow\mathbb{R}$ by $U(\varepsilon)=\tilde{V}_0(d^{\varepsilon}(\ell))$. Applying analogous reasoning to Lemma~\ref{brezislemma}, we have
\[
\lim_{\varepsilon\rightarrow 1} U'(\varepsilon)\propto(\phi(x)-\phi(y))\int_y^{x}\frac{(1+\beta)\frac{\phi''(z(1+\beta))}{\phi'(z(1+\beta))}-\frac{\phi''(z)}{\phi'(z)}}{(\phi'(z))^2}\,dz>0,
\]
which implies the existence of some $\bar{\varepsilon}<1$ such that $U'(\tilde{\varepsilon})>0$ for all $\tilde{\varepsilon}\in[\bar{\varepsilon},1)$. Hence,
\[
\tilde{V}_0(d^{1}(\ell))-\tilde{V}_0(d^{\bar{\varepsilon}}(\ell))=\int_{\bar{\varepsilon}}^{1}U'(\varepsilon)\,d\varepsilon>0,
\]
thus establishing $d^{1}(\ell)\geq_{C} d^{\bar{\varepsilon}}(\ell)\geq_{C} d^{iid}(\ell)$ but $d^{1}(\ell)\succ_0 d^{\bar{\varepsilon}}(\ell)$. Therefore, $\phi$ must satisfy IRRA.
\end{proof}

\noindent
\textbf{Proof of Corollary \ref{premiumcorollary}.}
\begin{proof}
For the correlated and iid lotteries we can 
write
\begin{align*}
f(\varepsilon) = & \, \frac{1}{2} \phi\left(x + \beta \phi^{-1} \left( \phi(x) \left(\frac{1}{2} + \frac{\varepsilon}{2} \right) + \phi(y) \left(\frac{1}{2} -   \frac{\varepsilon}{2} \right) \right)\right) \\
& + \frac{1}{2} \phi\left(y + \beta \phi^{-1} \left( \phi(x) \left(\frac{1}{2} -  \frac{\varepsilon}{2} \right) + \phi(y) \left(\frac{1}{2} +  \frac{\varepsilon}{2} \right) \right)\right)
= \tilde V_0(d^{corr}(\varepsilon)),
\end{align*}
and
\begin{align*}
g(\pi) = & \, \frac{1}{2} \phi\left(x(1 - \pi) + \beta \phi^{-1} \left( \phi(x(1 - \pi)) \frac{1}{2} + \phi(y(1 - \pi)) \frac{1}{2} \right)\right) \\
& + \frac{1}{2} \phi\left(y(1 - \pi) + \beta \phi^{-1} \left( \phi(x(1 - \pi)) \frac{1}{2} + \phi(y(1 - \pi)) \frac{1}{2} \right)\right)
= \tilde V_0(d^{iid}(\pi)).
\end{align*}
Because $\phi$ is strictly increasing and $c_0$ is the same in both
lotteries, the equation
$V_0(d^{corr}(\varepsilon)) = V_0(d^{iid}(\pi))$
is equivalent to
$\tilde V_0(d^{corr}(\varepsilon)) = \tilde V_0(d^{iid}(\pi))$,
that is, to $f(\varepsilon)=g(\pi)$.

Since $\phi\in \mathcal{C}^3$, we can take a second-order Taylor expansion
of $f$ around $\varepsilon=1$ and a first-order expansion of $g$ around
$\pi=0$:
\[
f(\varepsilon)
= f(1)+f'(1)(\varepsilon-1)+f''(1)(\varepsilon-1)^2
+ o\big((\varepsilon-1)^2\big),
\]
\[
g(\pi)
= g(0)+g'(0)\,\pi + o(\pi).
\]
Formally setting these expressions equal and neglecting higher-order terms
in $(\varepsilon-1)$ and $\pi$, we obtain the local approximation
\[
\pi(\varepsilon)
\approx
\frac{f(1)-g(0)}{g'(0)}
+ \frac{f'(1)}{g'(0)}(\varepsilon-1)
+ \frac{f''(1)}{g'(0)}(\varepsilon-1)^2.
\]

By the definitions above,
$f(1)=\tilde V_0(d^{corr}(1))$ and $g(0)=\tilde V_0(d^{iid}(0))$.
Moreover, $g'(0)<0$, since the value of the iid lottery decreases when
$\pi$ increases (all future consumption is scaled down by $(1-\pi)$).

The same calculations as in the proof of Theorem~\ref{theo1} yield
\[
f'(1)
= -\beta\frac{\big(\phi(x)-\phi(y)\big)}{4}
   \int_{y}^{x} \frac{\phi'(z(1+\beta))}{\phi'(z)}
   \frac{\big\{R_{\phi}(z(1+\beta)) - R_{\phi}(z)\big\}}{z}\,dz,
\]
and
\[
f''(1)
= \beta^2\frac{\big(\phi(x)-\phi(y)\big)^2}{8}
   \left\{\frac{\phi'(x(1+\beta))}{\phi'(x)} ER_{\phi}(x,x)
         + \frac{\phi'(y(1+\beta))}{\phi'(y)} ER_{\phi}(y,y)\right\}.
\]
Defining
\[
k_1 := -\frac{1}{g'(0)},\quad
k_2 := -\beta\frac{\big(\phi(x)-\phi(y)\big)}{4g'(0)},\quad
k_3 := -\beta^2\frac{\big(\phi(x)-\phi(y)\big)^2}{8g'(0)},
\]
which are all strictly positive since $g'(0)<0$, and substituting these
expressions into the approximation for $\pi(\varepsilon)$ gives
\eqref{eq:appropremium}, as claimed.
\end{proof}

 \noindent
\textbf{Proof of Theorem \ref{theo2}.}  \textit{Outline of the proof}.
Using a general result from  \cite{hardy1952inequalities} on certainty equivalents, I show that SCA implies that the certainty equivalent $\phi^{-1}\left(\mathbb{E}_m\phi\left(V_{t+1}\right)\right)$ is concave in utilities.\footnote{\cite{cerreia2011uncertainty}  provide a similar representation under the assumption that $\phi$ is strictly increasing and concave (see their Theorem 24). However, their result significantly differs from this one because they assume that $u(C)=(-\infty,\infty)$. This assumption is typically not satisfied in applications, such as the standard Epstein–Zin case.} This result allows us to utilize the Fenchel-Moreau duality theorem, revealing that the certainty equivalent can be represented dually as $\phi^{-1}\left(\mathbb{E}_m\phi\left(V_{t+1}\right)\right)=\min_{\ell}\mathbb{E}_{\ell} V_{t+1}+I_{\phi,u,\beta}^t(\ell\|m)$.

I introduce first some important notation: given a measurable space $(S,\Sigma)$, $c a(\Sigma)$ is the set of all countably additive elements of the set of charges $b a(\Sigma)$, while $c a_{+}(\Sigma)=c a(\Sigma) \cap b a_{+}(\Sigma)$ is its positive cone and $\Delta(\Sigma)$ is the set of countably additive probability measures. Given $p\in b a(\Sigma)$, let $b a(\Sigma, p)=\{v \in b a(\Sigma): B \in \Sigma$ and $p(B)=0$ implies $v(B)=0\}$. Observe that $b a(\Sigma, p)$ is isometrically isomorphic  (see \cite{dunford1958linear}, Theorem IV.8.16) to the dual of $L^{\infty}(p):=L^{\infty}(S,\Sigma,p)$ and $c a(\Sigma, p)=c a(\Sigma) \cap b a\left(\Sigma, p\right)$  is (isometrically isomorphic to) $L^1(p)$ (via the Radon-Nikodym derivative $\nu \mapsto \frac{d \nu}{ d p})$. 

Turning to the proof of Theorem \ref{theo2}, I first introduce  important notions related to quasi-arithmetic certainty equivalent functionals: given $p\in \Delta(\Sigma)$, let $M_{\phi,p}:L^{\infty}(p)\rightarrow\mathbb{R}\cup \{-\infty\}$ be defined by $$\phi^{-1}\left(\int\phi(\xi)dp\right)\text{ for every }\xi\in L^{\infty}(p),$$
assuming that $\phi:\mathbb{R}\rightarrow\mathbb{R}\cup\{-\infty\}$ is non-decreasing, concave, and upper semicontinuous. I provide an important result concerning the concave conjugate $M_{\phi,p}^*$ of the quasi-arithmetic mean $M_{\phi,p}$. Recall that  by the aforementioned isometry between the dual of $L^\infty(p)$  and $b a(\Sigma)$, the  concave conjugate $M_{\phi,p}^*$ can be seen as a mapping $M_{\phi,p}^*:ba(\Sigma,p)\rightarrow \mathbb{R}\cup\{\infty\}$ defined by $$M_{\phi,p}^*(q)=\inf_{\xi\in L^{\infty}(p)} \int \xi dq-M_{\phi,p}(\xi).$$
\begin{lemma}\label{maccheronzlemma}It holds that \( M_{\phi,p}^* \leq 0 \) and \( M_{\phi,p}^*(p) = 0 \) for every \( p \in \Delta(\Sigma) \). Moreover,  \( M_{\phi,p}^* \) is upper semicontinuous and concave when \( M_{\phi,p} \) is concave.  Finally, if \( \phi(x) = -\infty \) for any \( x < 0 \), \( \phi'(x) > 0 \), \( \phi''(x) < 0 \) for every \( x > 0 \), and \( p \in \Delta(\Sigma) \) has finite support, then the concave conjugate satisfies \( M_{\phi,p}^*(q) = -\infty \) whenever \( q \not\in \Delta(\Sigma) \cap \text{ca}(\Sigma, p) \).

\end{lemma}
\begin{proof}
Omitted.
\end{proof}


Denote by $L_{+}^\infty(p):=\{\xi\in L^\infty(p):\xi\geq 0\}$  the non-negative orthant of $L^{\infty}(p)$.
\begin{theorem}[See \cite{hardy1952inequalities} Theorem 106, \cite{chudziak2019weighted} or  \cite{gollier2001economics}]\label{concavechar}
Consider   $\phi:\mathbb{R}\rightarrow \mathbb{R}$ strictly increasing, strictly concave, and twice differentiable over $(0,\infty)$. Then  $M_{\phi,p}|L^{\infty}_{+}(p)$ is concave if and only if ${1}/{A_{\phi|(0,\infty)}}$ is concave.
\end{theorem}
\begin{proof}
If ${1}/{A_{\phi|(0,\infty)}}$ is concave,  it follows that by setting $L^\infty_{s,+}(p):=\{\xi\in L_{s,+}^\infty(p): \xi=\sum_{k=1}^n a_k \mathbf{1}_{A_k},(a_k)_{k=1}^n\in \mathbb{R}_{+}^n\}$,  one can apply  Theorem 1 and Theorem 5 in \cite{chudziak2019weighted} to show that $M_{\phi,p}|L_{s,+}^\infty(p)$ is concave. Concavity of $M_{\phi,p}|L_{+}^\infty(p)$ follows by the fact that $L_{s,+}^\infty(p)$ is dense in  $L_{+}^\infty(p)$. Conversely, if $M_{\phi,p}|L_{+}^\infty(p)$ is concave then $M_{\phi,p}|L_{s,+}^\infty(p)$ is also concave, which by Theorem 1 and Theorem 5 in \cite{chudziak2019weighted} implies that ${1}/{A_{\phi|(0,\infty)}}$ must be concave. 
\end{proof}
Thanks to Theorem \ref{concavechar}, we obtain the following powerful result, which shows that the conjunction of DARA and SCA on $\phi$ implies the concavity of the quasi-arithmetic mean $M_{\phi,p}|L^{\infty}_{+}(p)$. 
\begin{corollary}\label{genmarinaxmontru}Assume that $\phi\in \mathcal{C}^4$ is  concave, satisfies IRRA and UPI. Then   $R_{\phi}''\geq 0$ implies that  $M_{\phi,p}|L^{\infty}_{+}(p)$ is concave.   Conversely, if there exist $x\in (0,\infty)$ such that $R_{\phi}'(x)<0$  and $R_{\phi}''(x)<0$ then $M_{\phi,p}|L^{\infty}_{+}(p)$ is not concave. 
\end{corollary}
\begin{proof}
First observe that if $\phi$ satisfies UPI, then by DARA we have $A_{\phi}'\leq 0$. Further, it is immediately evident   that ${1}/{A_{\phi}}$ is concave whenever $$A_{\phi}''(x)A_{\phi}(x)\geq 2 (A_{\phi}'(x))^2,$$
for every $x\in (0,\infty)$. 
This condition is equivalent to  
\begin{equation}\label{hanoi}
xA_{\phi}''(x)\geq 2 x\frac{(A_{\phi}'(x))^2}{A_{\phi}(x)},
\end{equation}
for every $x\in (0,\infty)$.
Since $R_{\phi}'\geq 0$, we obtain that for every $x\in (0,\infty)$ it holds that $$A_{\phi}(x)\geq -x A_{\phi}'(x).$$
From this last condition we obtain that for every $x\in (0,\infty)$
\begin{equation}\label{taipei}
-2A_{\phi}'(x)\geq 2 x\frac{(A_{\phi}'(x))^2}{A_{\phi}(x)}.
\end{equation}
 Therefore since $$R_{\phi}''(x)=xA_{\phi}''(x)+2A_{\phi}'(x),$$
 if $R_{\phi}''\geq 0$ it follows that $xA_{\phi}''(x)\geq -2A_{\phi}'(x)$ which  by \eqref{taipei}  implies that \eqref{hanoi} is satisfied.
 Hence we conclude that if $\phi$ satisfies SCA then ${1}/{A_{\phi}}$ is concave. The result therefore follows by Theorem \ref{concavechar}.
 
 Conversely, if there exist  $x\in (0,\infty)$ such that  $R_{\phi}'(x),R_{\phi}''(x)<0$   we obtain that $-2A_{\phi}'(x)< 2 x\frac{(A_{\phi}'(x))^2}{A_{\phi}(x)}$ and $xA_{\phi}''(x)<-2A_{\phi}'(x)$  which implies that \eqref{hanoi} is violated, and so   by Theorem \ref{concavechar} we can conclude that  $M_{\phi,p}|L^{\infty}_{+}(p)$ is not concave.
\end{proof}
Now consider $\succeq$ with KP representation $(\phi,u,\beta)$. Without loss of generality, assume $u(C)=[0,\infty)$. I now show that letting \[\hat{\phi}(x)= \begin{cases} 
      \phi(x) &  x\geq 0 \\
      -\infty & x<0,
      \end{cases}
\]
then $M_{\hat{\phi},p}$ is concave if $\phi$ satisfies SCA. 
\begin{lemma}\label{kathechon}If $\phi:[0,\infty)\rightarrow\mathbb{R}$  satisfies SCA, then $M_{\hat{\phi},p}$ is  concave.
\end{lemma}
\begin{proof}The proof is a simple consequence of  Corollary \ref{genmarinaxmontru} and therefore is omitted.\end{proof}

It is important to observe that both EZ and HS preferences satisfy SCA.
\begin{corollary}\label{corollary} Assume that $\phi$ is given by $\phi(x)={x^{\lambda}}/{\lambda}$ for $0 \neq \lambda<1$ or $\phi(x)=-e^{-\frac{x}{\theta}}$ with $\theta>0$ for every $x\in\mathbb{R}_+$. Then $M_{\hat\phi,p}$ is concave. 
\end{corollary}
\begin{proof}
Immediate from Corollary \ref{genmarinaxmontru}, Theorem \ref{concavechar}, and Lemma \ref{kathechon}.
\end{proof}

It is now possible to deliver a proof of  Theorem \ref{theo2}.
\begin{proof}[Proof of Theorem \ref{theo2}]
Consider the utility functions $(V_t)_{t=0}^{T}$ from the KP representation $(\phi,u,\beta)$, observe that for every $m_t\in \Delta_b(D_t)$, where $\mathcal{D}_t$ is the Borel $\sigma$-algebra of $D_t$, since each $V_t:D_{t}\rightarrow\mathbb{R}$, $t=0,\ldots,T$ is upper semicontinuous in the weak$^*$ topology,  we have $V_{t}\in L_{+}^\infty(D_t,\mathcal{D}_t,m_t):=L_{+}^\infty(m_t)$. If $\phi$ satisfies SCA, then by Lemma \ref{kathechon} the function $M_{\hat{\phi},m_t}$ is concave for each $t=0,\ldots,T-1$. By applying the Fenchel-Moreau theorem (see \cite{phelps2009convex}, p. 42) and Lemma \ref{kathechon} it follows that
$$M_{\hat{\phi},m_t}(\xi^\prime)=\inf_{q \in\Delta(\mathcal{D}_t,m_t)}\mathbb{E}_q \xi -M_{\hat{\phi},m_t}^*(q)\quad\text{ for all }\xi\in L^\infty(m_t).$$

 Hence if for $t=0,\ldots, T-1$ we  set  \[I^t_{\phi,u,\beta}(\ell||m_t):=\begin{cases} -M_{\hat{\phi},m_t}^*(\ell), \\ 
+\infty  & \text{ otherwise},
 \end{cases}\]
 then one obtains that for every $(c,m_t)\in D_{t,s}$
 \begin{equation}\label{unilateral}
V_t(c,m_t)=u(c)+\beta\min_{\ell\ll m_t}\left\{\mathbb{E}_{\ell}V_{t+1}+I^t_{\phi,u,\beta}(\ell\|m_t)\right\},
\end{equation}
where the infimum is attained by Lemma \ref{maccheronzlemma} and because  $\{\ell\in \Delta_b(D_{t+1}):\ell\ll m_t\}$ is a compact  subset of  $\Delta_{b}(D_{t+1})$. Further, observe that by Lemma \ref{maccheronzlemma}, each is $I_{\phi,u,\beta}^t(\cdot\|m_t)$ is a convex statistical distance in the sense of \cite{liese1987convex}.
Now consider the  common parametrization of Epstein–Zin preferences used in asset pricing with ${1}/{(1-\rho)}>1$ and $\alpha<0$  (see \citealt{bansal2004risks}). In this case, one obtains (see  Section 5.2 in \citealt{frittelli1997certainty}) that  by setting  $q={\alpha}/{(\alpha-\rho)}$,
$$
I^t_{\phi,u,\beta}(\ell\|m_t)=\mathbb{E}_\ell{V_{t+1}}\left\{\left(\mathbb{E}_{m_t}\left[\left({\frac{d\ell}{d m_t}}\right)^q\right]\right)^{-\frac{1}{q}}-1\right\},
$$
so that upon noticing that the Rényi divergence is given for any $q>0$, $q\neq 1$   (see \citealt{van2014renyi}) by $$R_{q}(\ell\| m_t)=\frac{1}{q-1}\log\left(\mathbb{E}_{m_t}\left[\left({\frac{d\ell}{d {m_t}}}\right)^q\right]\right),$$
 we obtain that  for $t=0,\ldots,T-1$

\begin{equation}\label{eq:renyicost}
I_{\phi,u,\beta}(\ell\| m_t)=\mathbb{E}_\ell V_{t+1}\left[e^{\frac{1-q}{q} R_{q}(\ell \| {m_t})}-1\right].
\end{equation}
The Rényi divergence has applications in a variety of fields, including information theory, statistics, and machine learning (see \citealt{sason2022divergence} for a review). This result was already observed by \citeauthor{meyer2019generalized}, who showed that in the EZ case, the cost function can be expressed using Tsallis entropy (Rényi and Tsallis entropies are monotonic functions of each other; see, for example, \citealt{wong2022tsallis}). 

Finally, notice that if there exists $x\in (0,\infty)$ such that $R_{\phi}'(x)<0$  and $R_{\phi}''(x)<0$ then by Corollary  \ref{genmarinaxmontru} $M_{\phi,p}|L^{\infty}_{+}(p)$ is not concave, and so equation \eqref{unilateral} cannot hold for every  $(c,m_t)\in D_{t,s}$. 
\end{proof} 


\noindent
\textbf{Proof of Proposition \ref{prop:corrextension}.} It suffices to prove the following result.
\begin{lemma}\label{continuitylemma}
 If $(c,m_n)_{n=0}^\infty$ is a uniformly bounded sequence that converges to some $d\in D_0$, then $\lim_n V_0(c,m_n)=V_0(d)$
\end{lemma}
\begin{proof}
 Observe that
\begin{equation}\label{compactset} K\times \Delta_b(K),
\end{equation}
is a compact set whenever $K$ is compact (e.g., see Theorem 15.11 in \cite{guide2006infinite}). Further observe that $V_2(c)=u(c)$ when restricted to  a compact set $K$ is continuous and therefore bounded.  It follows that $V_1(c,m)=u(c)+\beta\phi^{-1}\mathbb{E}_m\left(\phi(V_2)\right)$ is continuous and therefore also bounded when restricted to the set  \eqref{compactset}. Finally, since the set in \eqref{compactset}   is compact, it follows that the set 
\begin{equation}\label{compactset2}
\Delta_b(K\times \Delta_b(K)),
\end{equation}
is also compact. Therefore, we can conclude that when restricted to the set in \eqref{compactset2} the function $V_0(c,m)=u(c)+\beta\phi^{-1}\mathbb{E}_m\left(\phi(V_1)\right)$ is continuous. 
\end{proof} 

\noindent
\textbf{Proof of Corollary \ref{corollaryperu}.}
\begin{proof}
 Observe that the equation $V_0(d^{gradual}(\varepsilon))=V_0(d^{early}(\pi))$ is equivalent to $\tilde{V}_0(d^{gradual}(\varepsilon))=\tilde{V}_0(d^{early}(\pi))$, and  so   if we set 
\begin{align*}
f(\varepsilon) &= \frac{1}{2} \phi\left(k + \beta \, \phi^{-1}\left(\phi(x)\left(\frac{1}{2} + \frac{\varepsilon}{2}\right) + \phi(y)\left(\frac{1}{2} - \frac{\varepsilon}{2}\right)\right)\right) \\
&\quad + \frac{1}{2}\phi\left(k + \beta \, \phi^{-1}\left(\phi(x)\left(\frac{1}{2} - \frac{\varepsilon}{2}\right) + \phi(y)\left(\frac{1}{2} + \frac{\varepsilon}{2}\right)\right)\right),
\end{align*}
and 
\begin{equation*}
g(\pi) = \frac{1}{2} \phi\left(k(1 - \pi) + \beta x(1 - \pi)\right)+ \frac{1}{2}\phi\left(k(1 - \pi) + \beta y(1-\pi)\right),
\end{equation*}
it becomes equivalent to $f(\varepsilon)=g(\pi)$. Since $\phi\in\mathcal{C}^2$, one can take the first order approximations of $f$ and $g$ so that
$f(\varepsilon)=f(1)+f'(1)(\varepsilon-1)+\mathrm{o}\left((\varepsilon-1)\right)$
and
$g(\pi)=g(0)+g'(0)\pi +\mathrm{o}\left(\pi\right)$.   Setting these expressions equal we obtain that for $\varepsilon\rightarrow 1$ $$\pi(\varepsilon)=\frac{f(1)-g(0)}{g'(0)}+\frac{f'(1)}{g'(0)}(\varepsilon-1)+\mathrm{o}\left((\varepsilon-1)\right).$$
 Furthermore, note that since $V_0(d^{gradual}(1))=V_0(d^{early}(0))$, it follows that  $f(1)=g(0)$. Moreover  $g'(0)<0$, since $\tilde{V}_0(d^{early}(\pi))$ obviously decreases with $\pi$. The same calculations as in the  proof of Proposition  \ref{peruchar} reveal   that $$f'(1)=\beta\frac{\phi(x)-\phi(y)}{4}\int_{y}^{x} \frac{\phi'(k+\beta z)}{\phi'(z)}ER_\phi(z,k)dz,$$
 which  setting $k_1=-\beta\frac{\phi(x)-\phi(y)}{4g'(0)}$  implies  the desired result.
 \end{proof}

\bibliography{ref}
\end{document}


\maketitle

\section*{Supplemental Appendix}
This supplemental material contains two parts.    Section  \ref{infinite} extends the main results to the case of an infinite  horizon.  Section \ref{sec:SCA} provides a behavioral characterization of strong correlation aversion.   Section \ref{proofs} provides a proof of Lemma \ref{brezislemma}.
\subsection{The case $T=\infty$}\label{infinite}
The theory presented thus far has focused on studying attitudes towards the correlation between consumption at two separate periods. However, it is also possible to consider more complex patterns of correlation, such as correlation between multiple periods. Here I consider the case of an infinite time horizon. As the consumption set $C=[0,\infty)=\mathbb{R}_{+}$ is identical to that of \cite{epstein1989substitution}, I follow their approach in introducing the set of temporal lotteries for the case of an infinite horizon, with specific reference to their discussion on pages 940-944. The only deviation in my approach is the use of $\Delta(X)$ to denote the set of Borel probabilities defined on a metric space $X$ (equipped with the weak$^*$ topology). For every  $b\geq 1$ and $l>0$, the sets of temporal lotteries $D(b;l)$ and $D(b)$ are defined in equations 2.3 and 2.5. Moreover, equations 2.2-2.11  define all the relevant objects. I also make use of their characterization of temporal lotteries in $D(b)$.
\begin{theorem}[Theorem 2.2 in \cite{epstein1989substitution}] For every $b\geq 1$ we have that
$$
D(b) \text { is homeomorphic to } C \times \hat{\Delta}(D(b)),
$$
where $$\hat{\Delta}(D(b)):=\left\{m \in {\Delta}(D(b)): f\left(m_2\right) \in \bigcup_{l>0} \Delta(Y(b ; l)), \quad m_2=P_2 m\right\}.$$
\end{theorem}
Because of this result, each $d\in D(b)$ can be identified with $(c,m)\in C\times \hat{\Delta}(D(b))$. Further, each $m\in \hat{\Delta}(D(b))$ can be equivalently identified with an element of $$ \hat{\Delta}(C\times \hat{\Delta}(D(b))).$$ Preferences are given by a weak order $\succeq$ over $D(b)$. 
The utility function $V:D(b)\rightarrow\mathbb{R}$ is called recursive if it satisfies the following equation for every $(c,m)\in C\times \hat{\Delta}\left(D(b)\right),$ 
\begin{equation}\label{recursiveutilityequation}
\quad V\left(c, m\right)=\left[c^\rho+\beta \phi^{-1}\left[\left(\mathbb{E}_m \phi\left(V\right)\right)\right]^\rho\right]^{1 / \rho}, \quad 0 < \rho<1, \quad 0<\beta<1,
\end{equation}
where $\phi:[0,\infty)\rightarrow\mathbb{R}$. The next result shows that \eqref{recursiveutilityequation}  always has a greatest and a smallest solution, thus making recursive utility well defined in this context.
\begin{theorem} 
Suppose that $\phi$ is concave, $\rho>0$ and $\beta b^\rho<1$. Then there exists a  $\bar{V},\ubar{V}:D(b)\rightarrow\mathbb{R}$ that satisfy  \eqref{recursiveutilityequation} and $\bar{V}\geq \ubar{V}$.
\end{theorem}
\begin{proof}
Denote by $S^{+}(D(b))$ the set of functions that map from $D(b)$ into positive real numbers. Let $h \in S^{+}(D(b))$ be  defined as in p. 963 of Appendix 3 in \cite{epstein1989substitution}. 
 Further, define $S_h^{+}(D(b))$ as follows
$$
S_h^{+}(D(b)) \equiv\left\{X \in S^{+}(D(b)):\|X\|_h \equiv \sup_{d\in D(b)} \frac{X(d)}{h(d)}<\infty\right\}.
$$
 Define $T: S_h^{+}(D(b)) \rightarrow S_h^{+}(D(b))$ by
$$T(X)=\left[c^\rho+\beta \phi^{-1}\left[\left(\mathbb{E}_m \phi\left(X\right)\right)\right]^\rho\right]^{1 / \rho}\quad\text{ for every }X\in S_h^{+}(D(b)).$$
Let $V^*$ be a continuous function such that
$$
\quad V^*\left(c_0, m\right)=\left[c^\rho+\beta \left[\mathbb{E}_m \left(V^*\right)\right]^\rho\right]^{1 / \rho}, \quad \rho>0, \quad 0<\beta<1,
$$
which exists uniquely by Theorem 3.1 in \cite{epstein1989substitution} since $\rho>0$ and $\beta b^\rho<1$.

 Let $T^0(V^*)=T(V^*)$ and $T^n(V^*)=T(T^{n-1}(V^*))$. By Jensen inequality $$\phi^{-1}\left(\mathbb{E}\phi(X)\right)\leq \mathbb{E} X \text{ for all } X\in S_h^{+}(D(b))\implies T(V^*)\leq V^*.$$ Further, it holds that $T (V^*) \geqslant 0$. By induction, one obtains that the sequence $(T^n(V^*))_{n=0}^\infty$ is non-increasing and bounded below. Therefore, we can define $V\in  S_h^{+}(D(b))$   as follows 
$$\bar{V}:= \lim_{n\rightarrow\infty} T^n V^*,$$

I now claim that $\bar{V}$ solves  \eqref{recursiveutilityequation}. Since
$$
T^n V^*\left(c_0, m\right)=\left[c^\rho+\beta \phi^{-1}\left[\left(\mathbb{E}_m \phi\left(T^{n-1}V^*\right)\right)\right]^\rho\right]^{1 / \rho}\quad\text{ for every }m\in D(b),
$$
the statement follows by the fact that
\begin{align*}
\lim_{n\rightarrow\infty}\phi^{-1}\left[\left(\mathbb{E}_m \phi\left(T^{n-1}V^*(m)\right)\right)\right]^\rho=\phi^{-1}\left[\left(\mathbb{E}_m \phi\left(\lim_{n\rightarrow\infty}T^{n-1}V^*(m)\right)\right)\right]^\rho\\=\phi^{-1}\left[\left(\mathbb{E}_m \phi\left(\bar{V}\right)\right)\right]^\rho.
\end{align*}
Hence, we can define $\ubar{V}\leq \bar{V}$ as follows
$$\ubar{V}=\lim _{n \rightarrow \infty} T^n(0).$$
Notice that $\ubar{V}$ is well defined because $\bar{V}$ is.
\end{proof}
Consider now a weak order $\succeq$ over $D(b)$. Say that $\succeq$ admits a KP representation $(\phi,\rho,\beta,V)$ if there exists $V:D(b)\rightarrow\mathbb{R}$ that satisfies \eqref{recursiveutilityequation} and  such that represents $\succeq$. For every  $d\in D(b)$ one can define the present equivalent $PE_{\succeq}(d)$   as the unique single period consumption level $c\in C$ such that $d\sim (c,\mathbf{0})$, where $\mathbf{0}\in D(b)$ is the temporal lottery that pays the constant zero level of consumption at every time period. Note that  $PE_{\succeq}(d)$ is well defined since $V(c,\mathbf{0})=c$.

 Now observe that every  $m\in \hat{\Delta}(C\times\hat{\Delta}(D(b)))$ and $\succeq$ with KP representation $(\phi,\rho,\beta)$ induce the probability $m_\succeq$ over ${\Delta}_b(C\times {\Delta}_b(C))$ defined as follows: 
$$m_\succeq(A\times B)=m(A\times B_\succeq)\quad\text{ for every closed } A\times B\subseteq C\times
\Delta_b(C),$$
where $B_\succeq=\{\ell\in \hat{\Delta}(D(b)):\ell_\succeq\in B\}$ and  $\ell_\succeq\in\Delta_b(C)$ is defined by $\ell_{\succeq}(A)=\ell(\{d\in D(b):PE_{\succeq}(d)\in A\})$.\footnote{The lottery $m_{\succeq}$ is well defined since preferences are continuous, $u(x)=x^\rho$ is strictly increasing and each $m\in \hat\Delta(D(b))$ has compact support.} In words, $m_{\succeq}$ describes the joint distribution between consumption at time $t+1$ and the continuation temporal lottery, where each temporal lottery is expressed in terms of one-period consumption. In this way, it is possible to extend the order $\geq_C$ and the correlation aversion axiom  as follows.
\begin{definition}[Correlation order with $T=\infty$] Fix  a weak order $\succeq$ over $D(b)$. Consider $d=(c,m),d'=(c,m')\in D(b)$. Say that $d$ is more correlated than $d'$, written $d\geq_C d'$,    if  $$m_\succeq,m_\succeq^\prime\in\Delta_s(C\times \Delta_s(C))\text{ and }  (c,m_\succeq)\geq_C (c,m_\succeq^\prime).$$ 
\end{definition}
Correlation aversion can  then be defined as  in the main text, where now $d^{iid}(\ell)=(c,m)$ denotes a temporal lottery such that $(c,m_\succeq)=d^{iid}(\ell)$ for some $\ell\in\Delta_s(C)$.
\begin{definition}[Correlation aversion with $T=\infty$]Say that $\succeq$ exhibits correlation aversion if and only if for every $l>0$ and  $d,d'\in D(b)$ $$d\geq_C d'\geq_C d^{iid}(\ell)\implies d^{iid}(\ell)\succeq d'\succeq d.$$
\end{definition}
The main results of the paper carry over in the same way. Notice  that due to the stationary setting, here there is a unique cost function $I_{(\phi,u,\beta)}(\cdot, \cdot)$.
\begin{theorem}\label{th:extinfinite}
Consider $\phi\in \mathcal{C}^3$ that is concave  and satisfies UPI. Then every $\succeq$ with KP representation $(\phi,\rho,\beta,V)$ exhibit correlation aversion if and only if $\phi$ satisfies IRRA. Further, if $\succeq$ admits a KP representation $(\phi,\rho,\beta,V)$ with $\phi \in \mathcal{C}^4$ that additionally satisfies SCA, then $\succeq$ admits the  representation for every $(c,m) \in C \times \big( \hat{\Delta}(D(b)) \cap \Delta_s(D(b)) \big)$
\[
V(c,m)=\left[ c^{\rho} + \beta \left( \min_{\ell \in \hat{\Delta}(D(b))} \big\{ \mathbb{E}_{\ell} V + I_{(\phi,u,\beta)}(\ell \| m) \big\} \right)^{\rho} \right]^{1/\rho},
\]
where $I_{(\phi,u,\beta)}(\cdot,\cdot)\colon \hat{\Delta}(D(b)) \times \hat{\Delta}(D(b)) \to [0,\infty]$ is a convex statistical distance.
\end{theorem}
\begin{proof}
The proof follows the same steps as the proof of Theorems \ref{theo1} and \ref{theo2}.
\end{proof}
To better understand the previous result, it is helpful to examine its implications. The following result demonstrates that if preferences satisfy this notion of correlation aversion, they will always prefer an iid lottery over a perfectly correlated one, where an iid lottery and a perfectly correlated lottery  are straightforward extensions of those considered in the main text.

Formally,  given \(\ell \in \Delta_s(C)\), consider the  perfectly correlated lottery \((c_0, m^{corr}(\ell))\), where \(m^{corr} \in \Delta(C^\infty)\) satisfies \(m^{corr}(c, c, \ldots) = \ell(c)\) for every $c\in C$, and the iid lottery \((c_0, m^{iid}(\ell))\), where 
\[
m^{iid}(\ell)(c, m^{iid}(\ell)) = \ell(c)\text{ for every }c\in C.
\]
These lotteries generalize the notions of iid lotteries and perfectly correlated lotteries from the case \(T =2\) to the case \(T = \infty\).

\begin{proposition}\label{prop:sa1}
Consider preferences \(\succeq\) with a KP representation \((\phi, \rho, \beta,\ubar{V})\) such that  $\phi$  satisfies IRRA,  UPI and $\phi\in \mathcal{C}^3$. 	Then $$(c_0, m^{iid}(\ell)) \succeq (c_0, m^{corr}(\ell)),$$ for every \(c_0 \in C\) and \(\ell \in \Delta_s(C)\).
\end{proposition}
\begin{proof}

Observe that  
\[
\left(\ubar{V}\left(c_0, m^{corr}(\ell)\right)\right)^\rho = c_0^\rho + \lim_{T \to \infty} \beta \phi^{-1} \left( \sum_{c \in \mathrm{supp} \ell} \ell(c) \phi \left( \sum_{t=0}^{T-1} \beta^t c^\rho \right) \right)
\]
and  
\[
\left(\ubar{V}\left(c_0, m^{iid}(\ell)\right)\right)^\rho = \lim_{T \to \infty} V_T\left(c_0, \ell\right),
\]
where \(V_0(c, \ell) = c^\rho\) and, recursively,  
\[
V_t(c, \ell) = c^\rho + \beta \phi^{-1} \left( \sum_{c' \in \mathrm{supp} \ell} \ell(c') \phi \big(V_{t-1}(c', \ell)\big) \right) \quad \text{for } t = 1, \ldots, T.
\]
Since the preferences \(\succeq\) satisfy correlation aversion,  Theorem \ref{th:extinfinite} implies that \(\phi\) satisfies IRRA. Additionally, by assumption, \(\phi\) satisfies  UPI and therefore  DARA (Proposition \ref{prop:upidara}). Hence, by   Proposition 6 and Theorem 12 in \cite{marinacci2010unique}, the functional $(x_i)_{i=1}^n\mapsto\phi^{-1}\left(\sum_{i=1}^n \phi\left(x_i\right) q_i\right)$ is  constant superadditive and subhomogeneous.  Therefore, by  repeatedly applying these results, we have that for every \(T \geq 2\) it holds
\[
V_T\left(c_0, \ell\right)-c_0^\rho\geq   \sum_{t=0}^{T-1} \beta^t \beta \phi^{-1} \left( \sum_{c \in \mathrm{supp} \ell} \ell(c) \phi(c^\rho) \right) \geq \beta \phi^{-1} \left( \sum_{c \in \mathrm{supp} \ell} \ell(c) \phi \left( \sum_{t=0}^{T-1} \beta^t c^\rho \right) \right).
\]

Consequently,   since
\[
\left(\ubar{V}\left(c_0, m^{iid}(\ell)\right)\right)^\rho = \lim_{T \to \infty} V_T\left(c_0, \ell\right) \geq \lim_{T \to \infty} \left(c_0^\rho + \beta \phi^{-1} \left( \sum_{c \in \mathrm{supp} \ell} \ell(c) \phi \left( \sum_{t=0}^{T-1} \beta^t c^\rho \right) \right) \right),
\]
and  
\[
\lim_{T \to \infty} \left(c_0^\rho + \beta \phi^{-1} \left( \sum_{c \in \mathrm{supp} \ell} \ell(c) \phi \left( \sum_{t=0}^{T-1} \beta^t c^\rho \right) \right) \right) = \left(\ubar{V}\left(c_0, m^{corr}(\ell)\right)\right)^\rho,
\]
it follows that  
\[
(c_0, m^{iid}(\ell)) \succeq (c_0, m^{corr}(\ell)),
\]
as desired.\end{proof}
\subsection{Axiomatic foundation of SCA}\label{sec:SCA}
Consider now the same setting of Section \ref{attcorr} of the main text. In order to provide an axiomatic foundation of SCA, I introduce the notion of \textit{Correlation Aversion Attenuation} (CAA) transformation. This integral operator modifies a given function $\phi$ to produce a new function, denoted by $CAA(\phi)$, which attenuates correlation aversion. Formally, the CAA transformation is a  nonlinear integral operator
\[
CAA: \mathcal{C}^4 \to \mathcal{C}^3,
\]
defined for every $\phi \in \mathcal{C}^4$ by:
\[
CAA(\phi)(x) = \int_1^x \exp\left(-\int_1^t \frac{R_\phi'(s)}{s} \, ds\right) dt.
\]
This integral transform  attenuates correlation aversion in that it ``flattens'' the index of relative risk aversion $R_{\phi}$, as I illustrate in the next example.
\begin{example}\label{ex:caat}
Given $\theta\in (0,1)\cup (1,\infty)$, let
\[
\phi(x) = -\exp\left(-\frac{x}{\theta}\right).
\]
Relative risk aversion is $
R_\phi(x) = \frac{x}{\theta}.
$ Applying the CAA operator, we obtain
\[
CAA(\phi)(x) = \frac{x^{1 - \frac{1}{\theta}}-1}{1 - \frac{1}{\theta}}.
\]
In this case, relative risk aversion is flat at the level  $R_{CAA(\phi)}(x) = \frac{1}{\theta}$. Applying the CAA operator once  again yields:
\[
CAA^2(\phi)(x) = x-1,
\]
which satisfies $R_{CAA^2(\phi)}(x) = 0.$ Hence, repeated applications of the CAA operator progressively attenuate correlation aversion by flattening the index of relative risk aversion.\demo
\end{example}

Consider preferences $\succeq$ that admit a KP representation $(\phi, u, \beta)$, where $\phi \in \mathcal{C}^4$.  Let $\succeq_{CAA}$ denote preferences with the KP representation $
(CAA(\phi), u, \beta)$.
\begin{definition}[Strong correlation aversion]Preferences $\succeq$ exhibit strong correlation aversion if  both $\succeq$ and $\succeq_{CAA}$ exhibit correlation aversion.
\end{definition}
Therefore, this notion of strong correlation aversion requires that  preferences exhibit correlation aversion even  after risk attitudes are adjusted to attenuate correlation aversion.
\begin{proposition}\label{suppapp:SCA}Assume that both $\phi$ and $CAA(\phi)$ satisfies UPI. Every preference relation $\succeq$ with KP representation $(\phi,u,\beta)$ exhibit strong correlation aversion if and only if    $\phi$ satisfies SCA.
\end{proposition}
\begin{proof}
Straightforward calculations show that $$R_{CAA(\phi)}(x)=R_{\phi}'(x).$$
Therefore since  $CAA(\phi) \in \mathcal{C}^3$ and $CAA(\phi)$ satisfies UPI, the result follows by Theorem \ref{theo1}.
\end{proof}
This  result shows that this behavioral notion of strong correlation aversion is effectively the behavioral counterpart of SCA.
\subsection{Proof of Lemma \ref{brezislemma}}\label{proofs}
Write the support of $m_1$ as $\{c_1,\ldots,c_N\}$ and $p_i=m_1(c_i)$ for every $i=1,\ldots,N$. Let $x_i=u(c_i)$ for $i=1,\ldots,N$ and $$U(\varepsilon)=\sum_{i=1}^N p_i\phi\left(x_i+\beta\phi^{-1}\left(\sum_{j=1}^Np_{ji}^\varepsilon\phi(x_{j})\right)\right)\quad \text{ for every }\varepsilon\in [0,1],$$
where   $$
p_{j i}^{\varepsilon}= \begin{cases}m_2^{\prime}\left(c_i \mid c_i\right)+\frac{\varepsilon \bar{\varepsilon}}{p_{\bar{i}}}, & (j, i)=(\bar{i}, \bar{i}), \\ m_2^{\prime}\left(c_j \mid c_i\right)-\frac{\varepsilon \bar{\varepsilon}}{p_{\bar{i}}}, & (j, i)=(\bar{j}, \bar{i}), \\ m_2^{\prime}\left(c_{\bar{j}} \mid c_{\bar{j}}\right)+\frac{\varepsilon \bar{\varepsilon}}{p_{\bar{j}}}, & (j, i)=(\bar{j}, \bar{j}), \\ m_2^{\prime}\left(c_{\bar{i}} \mid c_{\bar{j}}\right)-\frac{\varepsilon \bar{\varepsilon}}{p_{\bar{j}}}, & (j, i)=(\bar{i}, \bar{j}), \\ m_2^{\prime}\left(c_j \mid c_i\right), & \text { otherwise. }\end{cases}
$$
Clearly, the function $U$ defined in this manner is twice continuously differentiable and satisfies condition~(1) of the statement, where $\bar{\varepsilon}$ and $\bar{i},\bar{j}$ are the relevant IECIT parameters.

To prove point (2), observe that in this case we have that  for some $p,q\in (0,1)$, $k\in \phi(u(C))$ and $x,y\in u(C)$ with $x>y$
\begin{small}
\begin{align*}
\lim_{\varepsilon\rightarrow 0} U'(\varepsilon)= &\lim_{\varepsilon\rightarrow 0} \frac{\partial }{\partial \varepsilon} \Bigg[ p \phi\left(x+\beta \phi^{-1}\left(\phi(x)\left(p+\frac{\varepsilon}{p}\right)+\phi(y)\left(q-\frac{\varepsilon}{p}\right)+k\right)\right)+ \\
& q \phi\left(y+\beta \phi^{-1}\left(\phi(x)\left(p-\frac{\varepsilon}{q}\right)+\phi(y)\left(q+\frac{\varepsilon}{q} \right)+k\right)\right)\Bigg]\\
\leq  & \left(\phi(x)-\phi(y)\right) \lim_{\varepsilon\rightarrow 0}\Bigg[ \frac{\phi'\left(x+\beta \phi^{-1}\left(\phi(x)\left(p+\frac{\varepsilon}{p}\right)+\phi(y)\left(q-\frac{\varepsilon}{p}\right)+k\right)\right)}{\phi'\left(\phi^{-1}\left(\phi(x)\left(p+\frac{\varepsilon}{p}\right)+\phi(y)\left(q-\frac{\varepsilon}{p}\right)+k\right)\right)}- \\ & \frac{\phi'(y+\beta \phi^{-1}\left(\phi(x)\left(p-\frac{\varepsilon}{q}\right)+\phi(y)\left(q+\frac{\varepsilon}{q}\right)+k\right))}{\phi'\left(\phi^{-1}(\phi(x)\left(p-\frac{\varepsilon}{q}\right)+\phi(y)\left(q+\frac{\varepsilon}{q}\right)+k)\right)}\Bigg]\\
= & \left(\phi(x)-\phi(y)\right)\Bigg[ \frac{\phi'(x+\beta \phi^{-1}\left(\phi(x) p +\phi(y) q+k\right))}{\phi'\left(\phi^{-1}(\phi(x) p +\phi(y) q+k)\right)}-\\ & \frac{\phi'(y+\beta \phi^{-1}\left(\phi(x)p +\phi(y) q+k\right))}{\phi'\left(\phi^{-1}(\phi(x)p+\phi(y)q+k)\right)}\Bigg] \\
= &   \frac{(\phi(x)-\phi(y))}{\phi'\left(\phi^{-1}(\phi(x)p+\phi(y)q+k)\right)}\int_{y}^x \phi''\left(z+\beta\phi^{-1}\left(\phi(x)p+\phi(y)q+k\right)\right)dz   \leq 0,
\end{align*}
\end{small}
where the last inequality follows by the fact that $\phi$ is strictly increasing  and concave. 

Finally, to prove point (3), observe that the functions $$g_1(\varepsilon):=p_{\ubar i}\phi\left(x_i+\beta\phi^{-1}\left(\sum_{j=1}^N p_{j\ubar{i}}^\varepsilon\phi(x_{j})\right)\right),$$ and $$g_2(\varepsilon):=p_{\ubar{j}}\phi\left(x_i+\beta\phi^{-1}\left(\sum_{j=1}^N p_{j\ubar{j}}^\varepsilon\phi(x_{j})\right)\right),$$  are convex by Lemma \ref{rital} in the main text. Then we obtain
\begin{align*}
U''(\varepsilon)= & \frac{\partial^2 }{\partial \varepsilon^2} \Bigg[p_{\ubar i}\phi\left(x_i+\beta\phi^{-1}\left(\sum_{j=1}^N p_{j\ubar{i}}^\varepsilon\phi(x_{j})\right)\right)+p_{\ubar{j}}\phi\left(x_i+\beta\phi^{-1}\left(\sum_{j=1}^N p_{j\ubar{j}}^\varepsilon\phi(x_{j})\right)\right)\Bigg]\\ = & g_1''(\varepsilon)+ g_2''(\varepsilon)\geq 0,
\end{align*}
for every $\varepsilon\in(0,1)$ as desired. 

\bibliography{ref}